\documentclass[reprint,pra,notitlepage]{revtex4-1}
\usepackage{newtxtext}
\usepackage{newtxmath}
\let\lambda\lambdaup
\newcommand{\ud}{\mathrm{d}}
\usepackage{amsmath}
\usepackage{amssymb}
\usepackage{graphicx}
\usepackage{color}
\usepackage[colorlinks=true,linkcolor=blue,citecolor=blue,urlcolor=blue]{hyperref}
\graphicspath{{paperfig/}}

\allowdisplaybreaks
\begin{document}

\title{Non-power-law universality in one-dimensional quasicrystals}
\author{Attila Szab\'o}
\author{Ulrich Schneider}
\affiliation{Cavendish Laboratory, University of Cambridge, Cambridge CB3 0HE, United Kingdom}

\begin{abstract}
    We have investigated scaling properties of the Aubry--Andr\'e model and related one-dimensional quasiperiodic Hamiltonians near their localisation transitions.
    We find numerically that the scaling of characteristic energies near the ground state, usually captured by a single dynamical exponent, does not obey a power law relation. Instead, the scaling behaviour depends strongly on the correlation length in a manner governed by the continued fraction expansion of the irrational number $\beta$ describing incommensurability in the system. This dependence is, however, found to be universal between a range of models sharing the same value of $\beta$. For the Aubry--Andr\'e model, we explain this behaviour in terms of a discrete renormalisation group protocol which predicts rich critical behaviour.
    This result is complemented by studies of the expansion dynamics of a wave packet under the  Aubry--Andr\'e model at the critical point. Anomalous diffusion exponents are derived in terms of multifractal (R\'enyi) dimensions of the critical spectrum; non-power-law universality similar to that found in ground state dynamics is observed between a range of critical tight-binding Hamiltonians.
\end{abstract}

\maketitle

\section{Introduction}
\label{sec:intro}

Quasiperiodic structures, which are long-range ordered without being periodic, represent a rich and fascinating middle ground between ordinary periodic crystals and disordered systems. They were first discovered among aperiodic tilings of the plane, the best known of which is the fivefold symmetric Penrose tiling \cite{Berger1966,Penrose1974}. Interest in quasiperiodicity within the physics community was sparked by the discovery of quasicrystals by Shechtman \cite{Shechtman1985} and the equivalence between Landau levels on two-dimensional lattices and a one-dimensional quasiperiodic chain \cite{Harper1955,Azbel1964,Hofstadter1976}.
Recently, quasiperiodic structures became popular in ultracold atom experiments as a proxy for random potentials  in the study of disordered quantum gases, Bose glasses, and many-body localisation, as they can conveniently be realised by superimposing two incommensurate optical lattices \cite{Roati2008,Deissler2010,Sanchez2010,Schreiber2015,Luschen2017,Bordia2017}. 
Quasiperiodic tilings also lie at the heart of recent results in the study of quantum complexity, such as the proof of the undecidability of the spectral gap \cite{Cubitt2015}.

Quasiperiodicity gives rise to a range of unusual behaviour including critical spectra and multifractal eigenstates away from phase transitions \cite{Kohmoto1983,Ostlund1983,Kohmoto1986,You1991,Han1994,Liu2015} and localisation transitions at a finite modulation of the on-site potential \cite{Aubry1980,Han1994,Liu2015}. 
In this paper, we investigate localisation transitions of one-dimensional quasiperiodic systems, in particular the tight-binding Aubry--Andr\'e model, also known as the Harper model \cite{Aubry1980,Harper1955}:
\begin{equation}
    H = -J \sum_{j} \left( a_j^\dagger a_{j+1} + \mathrm{H.c.}\right) - J\lambda  \sum_{j} \cos(2\pi\beta j) a_j^\dagger a_j
    \label{eq:aubry}
\end{equation}
and related models. Here $\beta\not\in\mathbb{Q}$ and $\lambda$ are the incommensurate wave number and dimensionless amplitude of the on-site energy modulation, respectively, $J$ is the hopping matrix element,
and $a^\dagger_j$ is a bosonic creation operator on the $j$th lattice site. Since the integer part of $\beta$ is irrelevant, we assume $\beta<1$.
This model is known to undergo a localisation transition at $\lambda=2$ for any irrational value of $\beta$ \cite{Aubry1980,Suslov1982,Soukoulis1982,Jitomirskaya1999}: below this critical value, all eigenstates are extended while above it, they are exponentially localised. This is a consequence of Aubry duality: under the Fourier transform
\begin{equation}
    b_k = \frac1{\sqrt{\mathcal{N}}} \sum_n \exp\left(2\pi i\beta kn\right) a_n,
    \label{eq:aubryduality}
\end{equation}
(\ref{eq:aubry}) turns into another Aubry--Andr\'e Hamiltonian in momentum space with $\lambda$ changed to $4/\lambda$ and all energies rescaled by a factor of $\lambda/2$ \cite{Aubry1980}: $\lambda=2$ is the fixed point of this transformation.

It is well known that the spectra of one-dimensional quasiperiodic Hamiltonians are hierarchical \cite{Suslov1982,Wilkinson1984,Niu1986,Niu1990,Kohmoto1983,Kohmoto1986,Hatsugai1990,You1991,Azbel1979,Avila2009}, meaning they contain a hierarchy of progressively smaller gaps. In the case of tight-binding models, the spectrum is bounded and its entire structure is governed by the continued fraction expansion of the incommensurate ratio $\beta$ \cite{diophantine,Suslov1982,You1991},
\begin{align}
    \nonumber \beta &= \frac1{n_1+\beta_1} = \cfrac1{n_1+\cfrac1{n_2+\beta_2}} = \dots = \cfrac1{n_1+\cfrac1{n_2+\cfrac1\ddots}}\\*
    &=:[0;n_1,n_2,n_3,\dots],
    \label{eq:contfrac}
\end{align}
where $n_k$ are integers and the irrational residuals $\beta_k$ are between 0 and 1. 
The hierarchical spectrum of these Hamiltonians
can be constructed as the limiting case of periodic superlattices with increasing periods $N_k$  described by rational approximants of $\beta\approx M_k/N_k = [0;n_1,\dots,n_k]$, as  discussed in detail below. In going from the $k$th-order superlattice to the $(k+1)$st, each band of the spectrum is split into $n_{k+1}$ new ones \cite{Suslov1982}, see Fig.~\ref{fig:spectrum}. In this manner, the periods of these approximant superlattices, $N_k$, act as `microscopic length scales' of the problem: the structure of the spectrum and eigenstates of the Hamiltonian on length scales around each $N_k$ is controlled solely by the coefficient $n_k$. As a consequence, the spectrum is self-similar if and only if the continued fraction expansion of $\beta$ is periodic. Furthermore, its hierarchy is topologically protected under smooth deformations between different models sharing the same value of $\beta$ \cite{Dana2014,Thouless1982}.

In a continuous phase transition, the correlation length $\xi$ diverges at the transition point. In conventional  disordered or crystalline systems, the effect of microscopic structure becomes immaterial once $\xi$ is much larger than all microscopic scales of the system. Therefore, their behaviour near the phase transition is described by scale-invariant functions, that is, power laws \cite{Chaikin}. 
In quasiperiodic systems, however, such a scaling regime is never reached due to the increasingly large `microscopic' length scales $N_k$ discussed above. Instead, the behaviour of the system  is governed by scaling properties of the critical spectrum and eigenstates at length scales close to $\xi$ as it diverges, which in turn depends on the coefficients $n_k$.  While the connection between the structure of the spectrum and the length scales of the system has tacitly been known, its effects on phase transitions were not discussed, mostly because all numerical and most analytical studies focused on $\beta$'s of particularly simple continued fraction expansions, such as the golden mean $\phi^{-1}=(\sqrt{5}-1)/2=[0;\overline{1}]$ \cite{Kohmoto1986,Kohmoto1983,Ostlund1983,Cestari2010,Cestari2011}. (The overbar denotes a periodic continued fraction, \emph{e.g.,} $[0;1,\overline{2,3}] = [0;1,2,3,2,3,\dots]$.)

In this paper, we explore some consequences of this non-power-law critical behaviour on the localisation transition of the single-particle Aubry--Andr\'e model (\ref{eq:aubry}) using exact diagonalisation and renormalisation group arguments. In particular, we investigate the critical dynamics of the model for different values of $\beta$ and demonstrate that power-law behaviour emerges only when the continued fraction expansion of $\beta$ is periodic.

Section~\ref{sec:spectrum} reviews the origins of hierarchical spectra in quasiperiodic systems and presents a renormalisation group treatment of the Aubry--Andr\'e model based on Ref.~\onlinecite{Suslov1982}. In Sec.~\ref{sec:sff}, we discuss the scaling of energy scales near the ground state; Sec.~\ref{sec:multi} deals with fractal properties of the spectrum and quench dynamics at critical points. In both cases, we find  equivalent behaviour for different models sharing the same $\beta$. We understand this equivalence as a novel kind of universality, distinct from power-law thermodynamic universality, but similarly protected by symmetries of the underlying systems. Conclusions are presented in Sec.~\ref{sec:conclusion}.

\section{Structure of the spectrum and eigenstates}
\label{sec:spectrum}

\subsection{Structure of the critical spectrum}
\label{sec:spectrum:qual}

\begin{figure}
    \centering
    \includegraphics{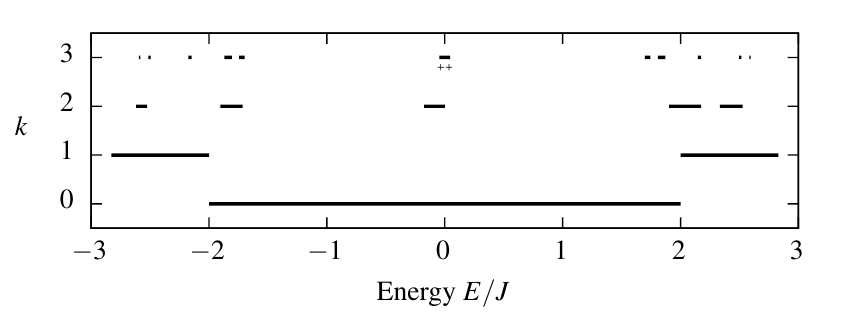}
    \caption{Spectrum of the Aubry--Andr\'e model for different rational $M_k/N_k\approx[0;\overline{2}]$ at the critical point $\lambda=2$. In each case, the spectrum consists of $N_k$ bands, most of which are accounted for by splitting the $N_{k-1}$ bands of the previous rational approximation into $n_k$ narrower ones. Some additional bands appear due to the slight changes to the approximation of $\beta$ \cite{Thouless1983,Tang1986,Hofstadter1976,Stinchcombe1987,Bell1989}. Double crosses denote a pair of bands with a very small gap, not resolved well in the plot.}
    \label{fig:spectrum}
\end{figure}

We consider a one-dimensional quasiperiodic tight-binding system characterised by the incommensurate ratio $\beta$. For simplicity, we assume that the continued fraction terms of $\beta$, $n_k$, are all very large; however, the qualitative structure of the spectrum remains the same for all $n_k\ge 2$  \cite{Hofstadter1976,Stinchcombe1987,Bell1989,Note11}\footnotetext[11]{The case of $n_k=1$ is special. It implies $\beta_{k-1}>1/2$ which can be replaced with $1-\beta_{k-1}$ without changing the resulting structure: the first continued fraction term of this number is, however, greater than 1.}. 
Now, as discussed in Sec.~\ref{sec:intro}, the structure of the spectrum can be described in terms of a sequence of periodic superlattices described by $ M_k/N_k=[0;n_1,\dots,n_k]$, which are the closest rational approximations of $\beta$ in the sense that \cite{diophantine}
\begin{equation*}
    |N_k\beta-M_k|<|N\beta-M|\quad \forall M,N\in\mathbb{Z}, 0<N<N_k.
\end{equation*}
At the first step of this protocol, $ M_1/N_1=1/n_1$: Bloch's theorem applies to the superlattice of period $n_1$, resulting in a spectrum consisting of $n_1$ subbands with continuous dispersion. At the next step, the period of the superlattice and thus the number of subbands is $N_2=n_1n_2+1\approx n_1n_2$ \footnote{Not all basis states of the system are accounted for in this approximation. This is compensated by the appearance of an additional $k$th order subband in the middle of each $(k-2)$nd order subband (see Fig.~\ref{fig:spectrum}), also described by the incommensurate ratio $\beta_k$ \cite{Hofstadter1976,Stinchcombe1987,Bell1989}. The approximation, however, provides a good description of states near the edges of the spectrum \cite{Suslov1982}.}. Since the approximation to $\beta$ changes very little, the spectrum is still dominated by the $n_1$ first-order bands, each now split into $n_2$ narrower subbands, see Fig.~\ref{fig:spectrum}. As further continued fraction terms are taken into account, more and more narrow subbands are formed, each time by splitting existing subbands into $n_k$ new ones.

\begin{figure}
    \centering
    \includegraphics{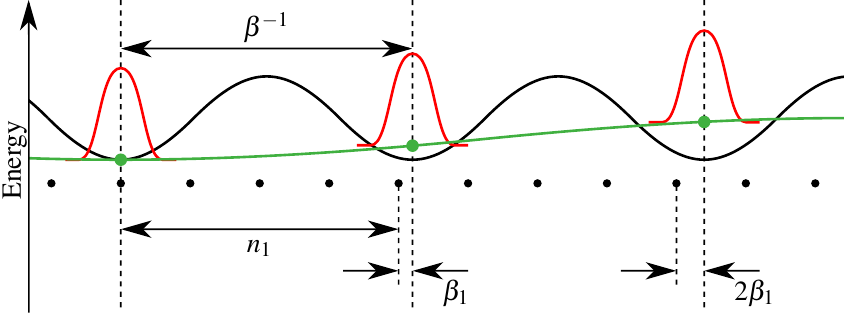}
    \caption{Cartoon of the renormalisation transformation of the lowest subband of the Aubry--Andr\'e model. 
    Black dots mark the lattice sites of the original tight-binding model; their quasiperiodic on-site potentials are indicated by the black line. In the first step of the transformation, every subband gives rise to one effective Wannier state (indicated in red) for each period of the on-site potential.
    For each subband, an effective tight-binding Hamiltonian  can now be defined by taking the centres of the corresponding Wannier states as new lattice sites. Due to the incommensurability of the on-site potential and the original lattice, the $j$th such Wannier state is shifted by $j\beta_1$ relative to the original lattice sites. As a result, the on-site energy of these new  sites will depend on $j$ in a quasiperiodic manner described by the incommensurate ratio $\beta_1$ (green line). The procedure is then repeated using the first-order Wannier states as lattice sites (green dots), introducing $\beta_2$, and so on indefinitely.}
    \label{fig:rg}
\end{figure}

The formation of this hierarchical structure can be understood in terms of a discrete renormalisation group procedure \cite{Suslov1982,Azbel1979,Niu1986,Niu1990,Wilkinson1984}. Creating first-order subbands can be taken as renormalising length scales by a factor of $\beta^{-1}\approx n_1$: the new `lattice sites' correspond to approximate Wannier states located at each minimum of the quasiperiodic modulation (see Fig.~\ref{fig:rg}). Since the modulation period is incommensurate to the lattice spacing, the $j$th renormalised lattice site will have a phase shift $2\pi\beta_1 j$ compared to the original lattice sites. This results in a quasiperiodic modulation of incommensurate ratio $\beta_1$ in the effective Hamiltonian of each subband. These Hamiltonians can now be renormalised by a factor of $\beta_1^{-1}\approx n_2$, giving rise to second-level subbands modulated with the new incommensurate ratio $\beta_2$: repeating such steps indefinitely constructs the entire spectrum. It can be shown \cite{Suslov1982} that for the Aubry--Andr\'e model with $n_k\gg1$, the renormalised on-site potential remains purely sinusoidal, that is, the effective Hamiltonian of each subband is of Aubry--Andr\'e form with incommensurate ratio $\beta_k$ at the $k$th step.

\subsection{Behaviour in the extended and localised regimes}

\begin{figure}
    \centering
    \includegraphics{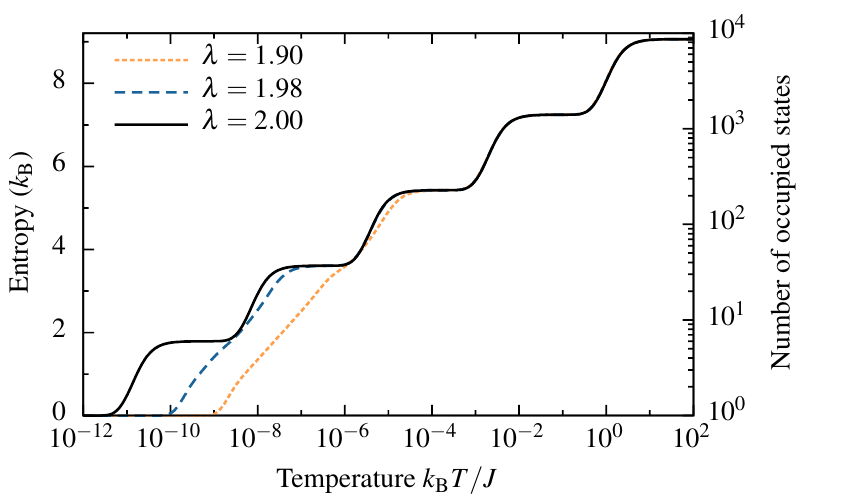}
    \caption{Thermal entropy per particle in the Aubry--Andr\'e model for $\beta = 1405/8658 \approx [0;\overline{6}]$ and different values of $\lambda$. 
    At criticality, the distinct sizes of gaps appearing at different renormalisation steps, and thus length scales $N_k$,  result in an infinite staircase structure: in a  finite portion of the model, this structure gets cut off at the narrowest band width of the system. 
    For $\lambda\neq2$,  the critical structure persists in the stairs corresponding to $N_k<\xi$. At lower energy scales, the system is effectively localised or extended, therefore, the temperature dependence of entropy is equivalent to that of an unmodulated tight-binding chain, $S(T)\simeq k_\mathrm{B} \log(T)/2$ at the lowest temperatures.}
    \label{fig:entropy}
\end{figure}

At $\lambda=2$ in the Aubry--Andr\'e model, the hierarchical structure of the spectrum described above is manifest at all energy scales and therefore all length scales. Away from the critical point, the correlation length $\xi$ of the system becomes finite. On this length scale, either the potential or the kinetic energy term of (\ref{eq:aubry}) becomes irrelevant, resulting in an either absolutely continuous spectrum and extended states for $\lambda<2$ or a dense point spectrum and exponentially localised states for $\lambda>2$. The crossover between the critical spectrum and the extended or localised spectra can be demonstrated using the thermal entropy of a single particle in a canonical ensemble:
\begin{align}
    S &= -k_\mathrm{B}\sum_i p_i \log p_i &
    p_i &= \frac{e^{-E_i/(k_\mathrm{B}T)}}{\sum_j e^{-E_j/(k_\mathrm{B}T)}} .
    \label{eq:entropy}
\end{align}
At temperature $T$, $\exp(S/k_\mathrm{B})$ is a measure of the number of states up to $\sim k_\mathrm{B}T$ above  the ground state. $S(T)$ is plotted for different values of $\lambda\le2$ in Fig.~\ref{fig:entropy}. At criticality, each renormalisation step defines a new energy scale resulting in an infinite staircase structure. For $\lambda\neq2$, such stairs persist down to energy scales corresponding to lengths on the order of $\xi$. Below this scale, the stairs smooth out and the scaling of entropy with temperature approaches that expected for an unmodulated tight-binding chain.

In the localised phase $\lambda>2$, $\xi$ is normally identified with the localisation length of the wave function envelope which can be calculated without detailed analysis of the wave function \cite{Thouless1972}: in the Aubry--Andr\'e case \cite{Aubry1980},
\begin{subequations}
\begin{equation}
\xi (\lambda>2) = \left(\log\frac\lambda2\right)^{-1}
\label{eq:corrlen1}
\end{equation}
for all eigenstates and all values of $\beta$. Due to Aubry duality, the structure of the Aubry--Andr\'e spectrum for modulation amplitudes $\lambda$ and $4/\lambda$ is identical save for an overall rescaling \cite{Aubry1980}. This implies that the length scale $\xi$ where the crossover happens in the two cases is the same, giving
\begin{equation}
\xi (\lambda) = \left|\log\frac\lambda2\right|^{-1}.
\label{eq:corrlen}
\end{equation}
\end{subequations}

\begin{figure}
    \centering
    \includegraphics{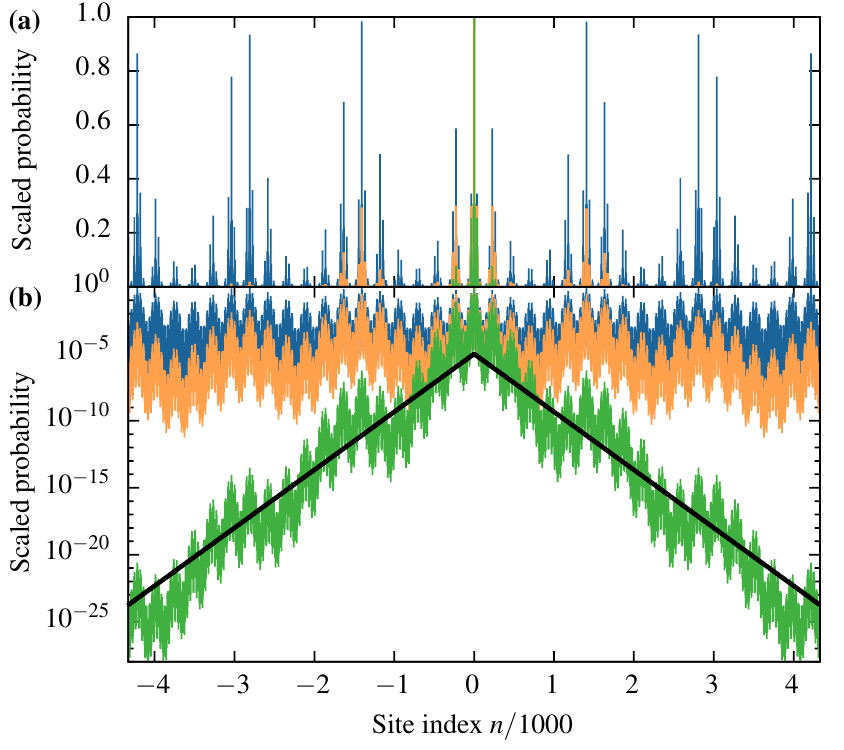}
    \caption{Scaled ground state probability distribution $|\psi(n)|^2/|\psi(0)|^2$ in the Aubry--Andr\'e model for $\beta = 1405/8658 \approx [0;\overline{6}]$ at the critical point ($\lambda=2$; orange), and for $\xi\approx 200$ in the extended ($\lambda=1.99$; blue) and the localised ($\lambda=2.01$; green) phases on linear (a) and logarithmic (b) scales. On length scales shorter than $\xi$, all wave functions appear similar; on larger scales, the coarse-grained density distribution of the extended state becomes uniform, while the localised wave function is dominated by exponential decay $\propto e^{-2|n|/\xi}$ (straight lines).}
    \label{fig:states}
\end{figure}

The crossover between critical and extended or localised behaviour is also manifest in the structure of the wave functions. At $\lambda=2$, nontrivial structure appears at all length scales: away from criticality, this structure is only manifest up to length scales $\approx\xi$ beyond which the density distribution is either dominated by exponential decay or becomes uniform. This is demonstrated for the ground state in Fig.~\ref{fig:states} which also confirms the localisation length given by (\ref{eq:corrlen1}).

\subsection{Analytic scaling theory}
\label{sec:spectrum:analytic}

In this section, we present a full renormalisation group treatment of the Aubry--Andr\'e model based on Ref.~\cite{Suslov1982}. This treatment becomes exact in the limit when all continued fraction terms of $\beta$ are large, that is, $\beta,\beta_1,\dots\ll1$.

As in Sec.~\ref{sec:spectrum:qual}, we start by approximating $\beta$ with $1/n_1$, that is, we consider the following periodic Hamiltonian:
\begin{align}
    H = -&J \sum_{j} \left( a_j^\dagger a_{j+1} + \mathrm{H.c.}\right) \nonumber\\
    -& J\lambda  \sum_{j} \cos\left(\frac{2\pi}{n_1} (j-\phi)\right) a_j^\dagger a_j,
    \label{eq:ham_rational}
\end{align}
where $\phi$ is a well-defined global spatial offset. 
The spectrum of the resulting periodic lattice splits into $n_1$  subbands each of which gives rise to  Wannier states with a spacing of $n_1$ lattice sites (red in Fig.~\ref{fig:rg}). Since $n_1$ is large, beyond-nearest-neighbour couplings between these new states are vanishingly small, and thus each subband can be well described by a new tight-binding model with dispersion
\begin{equation}
    E(\kappa,\phi) = E_0(\phi) - 2J'(\phi) \cos \kappa,
    \label{eq:dispersion1}
\end{equation}
where $J'$ is the hopping between two neighbouring Wannier states, $\kappa$ is the renormalised quasimomentum, and $E_0$ is the mean energy of the subband. In principle, both $E_0$ and $J'$ depend on the phase $\phi$ in (\ref{eq:ham_rational}). As we will discuss below, the variations of $E_0$ are on the order of $J'$, which is exponentially small. Similarly, for large $n_1$, the variations of $J'$ are exponentially smaller than $J'$ itself, thus they can safely be ignored in any effective theory.

We now determine the dependence of $E_0$ on $\phi$ in this periodic approximation by applying the Aubry duality transformation (\ref{eq:aubryduality}). It is important to note that, as $\beta$ is rational, (\ref{eq:aubryduality}) only generates $n_1$ distinct reciprocal space modes. In order to make the transformation unitary, (\ref{eq:ham_rational}) is replaced by a Hamiltonian acting on $n_1$ lattice sites with twisted periodic boundary conditions:
\begin{align}
    H = -&J\sum_{j=1}^{n_1} \left(a_j^\dagger a_{j-1} e^{i\kappa/n_1}+ \mathrm{H.c.}\right)\nonumber \\*
    - &J\lambda  \sum_{j=1}^{n_1} \cos\left(\frac{2\pi}{n_1} (j-\phi)\right) a_j^\dagger a_j,
    \label{eq:ham_rational_twisted}
\end{align}
where $a_0 = a_{n_1}$. Applying the duality transformation (\ref{eq:aubryduality}) to (\ref{eq:ham_rational_twisted}), it becomes
\begin{align}
    H = -\frac{J\lambda}2 &\bigg[\sum_{\ell=1}^{n_1} \left(b^\dagger_\ell b_{\ell-1} e^{2\pi i\phi/n_1} + \mathrm{h.c.}\right) \nonumber\\*
    &+\frac4\lambda\sum_{\ell=1}^{n_1} \cos\left(\frac{2\pi}{n_1}\Big(\ell-\frac{\kappa}{2\pi}\Big)\right) b^\dagger_\ell b_\ell \bigg].
\end{align}
That is, the duality transformation exchanges the quasimomentum $\kappa$ and the offset $\phi$. By (\ref{eq:dispersion1}), the energy eigenvalue of the dual Hamiltonian depends on its quasimomentum $2\pi\phi$ as a simple cosine the amplitude of which is taken independent of $\kappa$. Therefore, $E_0$ also has a cosine dependence on $\phi$:
\begin{align}
    E_0(\phi) &= E_0 - (J\lambda)' \cos(2\pi\phi)
    \label{eq:dispersion_phi}\\*
    (J\lambda)' &= \frac\lambda2\times 2J'_\mathrm{dual}.
    \label{eq:lambdat1}
\end{align}
Combining (\ref{eq:dispersion1}) and (\ref{eq:dispersion_phi}), the dispersion relation of the periodic approximation is finally given by
\begin{align}
    E(\kappa,\phi) &= E_0 -2J'\cos \kappa -(J\lambda )' \cos (2\pi\phi).
    \label{eq:dispersion2}
\end{align}

In the quasiperiodic system however, $1/\beta$ differs from $n_1$ by a small irrational number $\beta_1$, therefore, the $j$th minimum of the potential is shifted $\phi_j=j\beta_1$ away from a lattice site (see Fig.~\ref{fig:rg}). Equation~\ref{eq:dispersion2} is thus not exact, but  as $\beta_1$ is assumed to be small, $\phi_j$ changes slowly. Therefore, (\ref{eq:dispersion2}) can be used as an effective Hamiltonian for the new Wannier states of separation $1/\beta$. That is, upon rescaling by $1/\beta$, the resulting model is described by the Hamiltonian
\begin{equation}
    H =  -2J'\cos \hat{p}' -(J\lambda)' \cos (2\pi\beta_1 \hat{x}'),
    \label{eq:renorm}
\end{equation}
an Aubry--Andr\'e model of parameter $\beta_1$ with renormalised potential and hopping terms. The same procedure can then be repeated with step sizes $1/\beta_k$ to obtain a renormalisation group treatment of the full spectrum.

The terms $J', (J\lambda)'$ entering (\ref{eq:renorm}) may be estimated numerically from the scaling of bandwidths over a single step of the procedure. In the limit of large $n_k$, the scaling of $J$ for states sufficiently far from $E=0$ can be calculated analytically using the WKB approximation (see Appendix~\ref{app:wkb}). These calculations show that the renormalisation of the potential-to-hopping ratio $\lambda$ does not depend on energy (and hence the place of the subband in the spectrum), and is given by (see Appendix~\ref{app:lambda})
\begin{equation}
    \lambda' = 2 \left(\frac\lambda2\right)^{1/\beta}.
\end{equation}
Iterating this procedure on the emerging quasiperiodic lattices gives the effective amplitude $\lambda^{(k)}$ on length scale $\approx N_k$ as
\begin{equation}
    \lambda^{(k)} = 2 \left(\frac\lambda2\right)^{1/\beta\dots\beta_{k-1}} \simeq 2 \left(\frac\lambda2\right)^{N_k}.
\end{equation}
For $\lambda<2$, $\lambda > \lambda' > \lambda'' >\dots$: the RG procedure tends to $\lambda^{(k)}=0$, that is, the quasiperiodic modulation becomes irrelevant and hence all eigenstates are extended. On the other hand, if $\lambda>2$, $\lambda$ increases upon renormalisation: the system flows to $\lambda^{(k)}\to\infty$ where hopping is irrelevant, and all eigenstates are localised. The critical exponent of the reduced tuning parameter $g = \log(\lambda/2)$ is $\nu=1$: indeed, according to (\ref{eq:corrlen}), $|g|=\xi^{-1}$.

\section{Critical scaling near the superfluid--insulator transition}
\label{sec:sff}

We performed exact diagonalisation on the single-particle Aubry--Andr\'e Hamiltonian (\ref{eq:aubry}) and extrapolated the behaviour of the truly incommensurate model from the sequence of rational approximations $M_k/N_k$ of $\beta$, all implemented with periodic boundary conditions.

The key quantity we considered was the curvature of the lowest band,
\begin{equation}
    \Gamma = \frac1{2Ja_0^2}\left.\frac{\ud^2 \varepsilon(k)}{\ud k^2}\right|_{k=0} = \frac1J \frac{\hbar^2}{m_\mathrm{eff}a_0^2},
    \label{eq:sff0}
\end{equation}
where $m_\mathrm{eff}$ is the effective mass of particles near the bottom of the band and $a_0$ is the lattice spacing. The normalisation is chosen such that $\Gamma$ for an unmodulated tight-binding chain is unity. In an extended phase, the motion of a single particle becomes ballistic beyond a length scale, therefore, its effective mass tends to a finite value in the limit of an infinite system. Bands of a localised model, however, become completely flat, resulting in an infinite effective mass and thus zero curvature. As a consequence, the limit $\lim_{N\to\infty} \Gamma_{N}$, where $N$ is the period of the lattice,  can be used as an order parameter in a quantum localisation transition. We approximate the second derivative using the energy difference over a finite segment of the lowest band:
\begin{equation}
    \Gamma = \frac1J\frac{E_\Theta-E_0}{(\Theta/N)^2},
    \label{eq:sff}
\end{equation}
where $E_\Theta$ and $E_0$ are the ground state energies of the system in periodic boundary conditions twisted by $\Theta$ or without twist, respectively. Equations~\ref{eq:sff0} and~\ref{eq:sff} are equivalent for $\Theta\to0$, but in practice, $\Gamma$ remains essentially unchanged for significant fractions of $\pi$. In this paper, $\Theta=\pi/20$ was normally used.
In interacting many-particle systems, the appropriate generalisation of $\Gamma$ gives the superfluid fraction or superfluid stiffness, which is widely used to analyse superfluid--insulator transitions \cite{Lieb2002,Roth2003,Cestari2010}.

\begin{figure}
    \centering
    \includegraphics{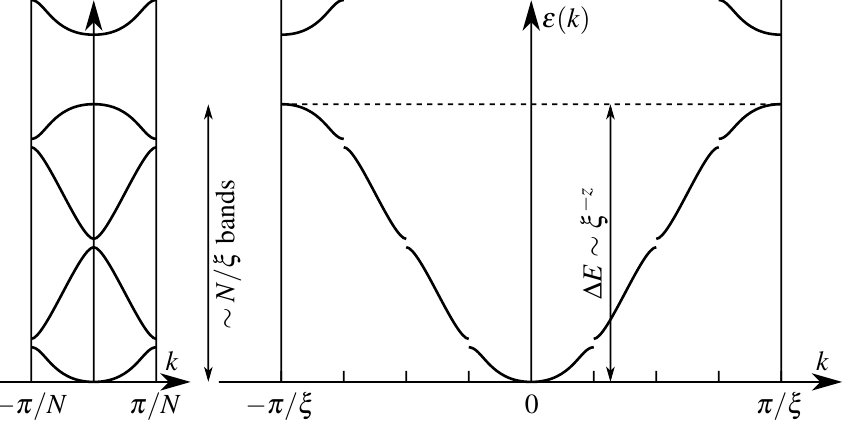}
    \caption{Cartoon of the extended lowest band of a periodic tight-binding model of period $N\gg\xi$. On length scales $N_k$ above the correlation length, any effective potential $\lambda^{(k)}$ is very small, resulting only in narrow avoided crossings. That is, the band structure is similar to an uninterrupted band of a model of period $\sim\xi$, as shown in the extended Brillouin zone on the right. The effective lowest band governing quantum critical dynamics is thus $\sim2\pi/\xi$ wide in momentum space and its width $\Delta E$ tends to a constant value as $N\gg\xi$: this can be used to estimate $\Gamma$ resulting in the scaling relation (\ref{eq:sff_scaling}).}
    \label{fig:folding}
\end{figure}

We note that the curvature of a band is related to its width $\Delta E$  and  therefore can be used to extract the scaling properties of the bandwidth; in a homogeneous or crystalline system, this scaling is governed by the dynamical exponent $z$:
\begin{equation}
    \Delta E \sim \xi^{-z}.
    \label{eq:zdef:energy}
\end{equation}
To elucidate this connection,  the typical band structure in the extended phase is sketched in Fig.~\ref{fig:folding}. 
On length scales $N_k\gg\xi$, the effective potential is irrelevant compared to the effective hopping (that is, the renormalised $\lambda^{(k)}\ll1$) and the spectrum of any periodic approximation with $N\gg\xi$ becomes similar to the spectrum for $N\approx\xi$: the \emph{effective} lowest band is folded up, largely conserving the continuity of the spectrum. In particular, the small gaps introduced by the remaining weak potential do not affect the curvature at $k=0$. That is, regardless of the period $N_k\gg\xi$ of the lattice, the lowest dynamical band is $\sim\pi/\xi$ wide in $k$-space. Approximating its dispersion by
\begin{equation*}
    \varepsilon(k) \sim -\frac{\Delta E}{2} \, \cos(\xi k),
\end{equation*}
the band curvature follows as
\begin{equation}
    \Gamma \sim \frac{\Delta E}{(1/\xi)^2} \sim \xi^{2-z} \sim |\lambda-2|^{\nu(z-2)}
    \label{eq:sff_scaling}
\end{equation}
by the definition of $z$ and $\nu$: note that $\nu=1$ for all $\beta$ in the Aubry--Andr\'e model (\emph{cf.}~Eq.~\ref{eq:corrlen}). We note that the scaling behaviour of the many-particle superfluid fraction is also given by (\ref{eq:sff_scaling}) \cite{Fisher1989}, as expected given its relation to $\Gamma$.

\subsection{Results for the Aubry--Andr\'e model}
\label{sec:sff:aubry}

\begin{figure}
    \centering
    \includegraphics{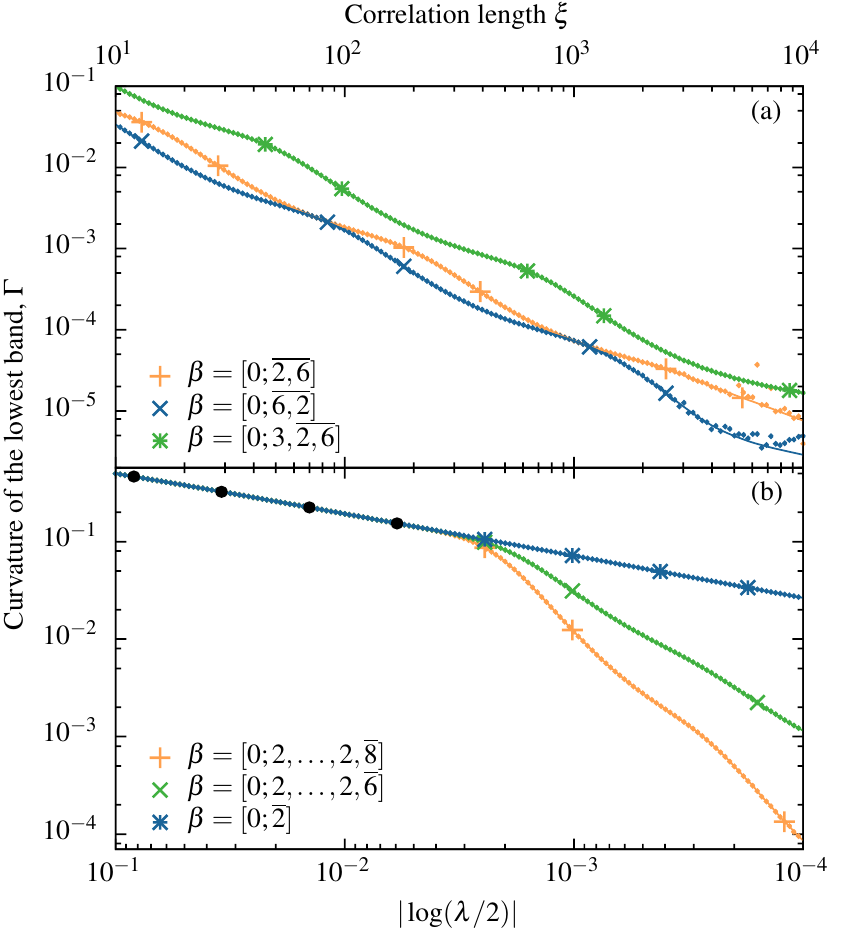}
    \caption{Curvature of the lowest band, $\Gamma$, as a function of the reduced tuning parameter $g=\log(\lambda/2)=\xi^{-1}$ for several incommensurate ratios $\beta$. Small dots indicate all computed data points, large symbols appear at $g=1/N_k$. Smoothing splines (thin solid lines) were added to two data sets to guide the eye. The period of the simulated superlattice for each curve is much greater than $\xi_\mathrm{max} = 10^4$. 
    (a) The continued fraction expansions of the $\beta$ are identical except for the first few terms. The sequences $\{\beta_k\}$ governing the fine structure of the spectrum are hence the same up to a shift, resulting in identical line shapes and effective critical exponents. The marks appear on the same parts of this line shape, indicating that the deviations from power law behaviour are caused by the hierarchical structure of the spectrum. 
    (b) The three values of $\beta$ differ by less than a part in $10^6$; however, their continued fraction expansions diverge after the eighth term ($N_8=985$), resulting in identical scaling up to $|g|\gtrsim1/985$ (black dots) followed by markedly different critical behaviour for $|g|\ll1/985$.}
    \label{fig:scaling}
\end{figure}

The curvature of the lowest band was calculated for the Aubry--Andr\'e model (\ref{eq:aubry}) near $\lambda=2$ for several different incommensurate ratios and plotted in Fig.~\ref{fig:scaling}. The rational approximations to $\beta$ were always chosen  such  that the period of the resulting superlattice was much larger than the longest correlation length considered, $\xi_\mathrm{max}=10^4$. In contrast to homogeneous systems, the order parameter never follows a power law, even when $\xi$ is on the order of thousands of lattice sites. This contradicts the conventional notion of a `scaling regime' where the only relevant length scale is the correlation length, resulting in power law behaviour \cite{Chaikin}. 

The origin of this discrepancy is the emergence of the arbitrarily large `microscopic' length scales $N_k$ discussed in Secs.~\ref{sec:intro} and~\ref{sec:spectrum:qual}. Consider a near-critical Hamiltonian with extended eigenstates of correlation length $\xi\approx N_k$: broadly speaking, its spectrum displays the first $k$ levels of the hierarchical critical spectrum, but further ones are not resolved and thus have no effect on $\Gamma$  (\emph{cf.}~Figs.~\ref{fig:entropy} and~\ref{fig:folding}). As a result, its critical scaling  at $\xi\approx N_k$ depends on the $k$th step of the renormalisation protocol of Sec.~\ref{sec:spectrum} which is in turn governed by $\beta_k$. In particular, the slope of the log-log plot in Fig.~\ref{fig:scaling} is determined by the local dynamical exponent $z_k = z(\beta_k)$ defined by
\begin{equation}
    \frac{\Delta E(N_{k+1})}{\Delta E(N_k)} =\left(\frac{N_{k+1}}{N_k}\right)^{-z_k}\cong \beta_k^{z_k}.
    \label{eq:z_local}
\end{equation}

In Fig.~\ref{fig:scaling}(a), the continued fraction expansions of all values of $\beta$ become periodic with identical periods; this implies that the sequence $\{\beta_k\}$ and thus the scaling behaviour is identical from a point on. This is manifest in the identical but shifted curves in the plot; the difference in overall scaling stems from the different initial terms in the continued fraction expansion which result in different $N_k$'s corresponding to the same $\beta_k$'s. 

In Fig.~\ref{fig:scaling}(b), the values of $\beta$ are very close to each other, and so their continued fraction expansions  start with the same terms. Since the first few $\beta_k$ differ by very little, the critical scaling is almost identical for relatively small $\xi$: this changes noticeably as further terms in the continued fraction expansions become different, giving rise to completely different scalings. This behaviour demonstrates that while the structure of quasiperiodic systems described by only slightly different incommensurate ratios may be very different on sufficiently long length scales, such differences are immaterial in short samples. Such unpredictability of the large-scale behaviour of quasiperiodic systems also plays a key role in quantum complexity theory \cite{Cubitt2015}.

We have thus found that the existence of `microscopic' structure on all length scales prevents the formation of a conventional scaling regime where exact power-law scaling relations such as (\ref{eq:sff_scaling}) would hold. For $\beta$'s with periodic continued fraction expansions, however, the sequence  $\{\beta_k\}$ itself is periodic and so the scaling behaviour repeats itself on arbitrarily long length scales.  In this case, one can combine all renormalisation steps in one period into a discrete RG protocol where all steps are identical. For such RG schemes, it is common to find a power-law behaviour \emph{on average}, with log-periodic oscillations around it \cite{Karevski1996,Nauenberg1975,Bessis1983,Doucot1986,Derrida1984}: indeed, we observe such oscillations in Fig.~\ref{fig:scaling}(a). Nonetheless, a single period of these oscillations may contain  an arbitrarily complex pattern of the RG steps defined in Sec.~\ref{sec:spectrum:analytic}, and so a description of the critical behaviour in terms of log-periodic oscillations is not generally practical.

The continued fraction expansions of almost all irrational numbers are, however, not periodic. For these numbers, the RG protocol cannot be described in terms of a single, if complex, step, resulting in a situation more complicated than the log-periodic oscillations discussed above. In particular, there is no way to sensibly define single critical exponents for the Aubry--Andr\'e model for these values of $\beta$. The critical behaviour is only appropriately described by the detailed dependence of observables on the length scale, an example of which is the set of local dynamical exponents (\ref{eq:z_local}). Using the analytic RG procedure discussed in Sec.~\ref{sec:spectrum:analytic}, $z_k$ can be calculated for $\beta_k\ll1$ (see Appendix~\ref{app:z}). To leading order,
\begin{equation}
    z_k\approx 1.166\ \frac{\beta_k^{-1}}{\log(\beta_k^{-1})},
    \label{eq:z_anal}
\end{equation}
meaning that $z_k\to\infty$ as $\beta_k\to0$. Therefore, for an incommensurate ratio $\beta=[0;n_1,n_2,\dots]$ with $n_{k_1}<n_{k_2}$ for all $K<k_1<k_2$ for some $K$, the conventional definition of the dynamical exponent,
\begin{equation}
    z = \lim_{\xi\to\infty} \frac{\log \Delta E}{\log\xi},
\end{equation}
diverges: we note that these numbers form a dense, uncountable subset of $[0,1]$. This marks a completely novel critical behaviour, one not even approximated by power laws.

\subsection{Ground state universality of quasiperiodic models}
\label{sec:sff:universal}

\begin{figure}
    \centering
    \includegraphics{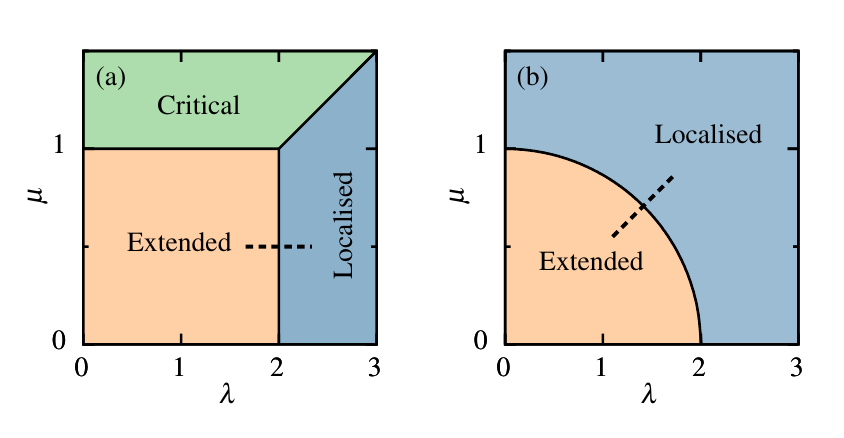}
    \caption{Phase diagram of the generalised Aubry--Andr\'e model (\ref{eq:aubry_gen}) for $\phi=0$ (a) and $\phi=\pi/2$ (b) \cite{Liu2015}. For all values of $\phi$, a localisation phase transition line appears in the $(\lambda,\mu)$ plane; additionally,  a  critical phase dominated by $\mu$ appears if $\phi=0$ precisely. The dashed lines show the paths (\ref{eq:gen_instance}) along which localisation transitions were considered in Sec.~\ref{sec:sff:universal}.}
    \label{fig:phases}
\end{figure}

In addition to the Aubry--Andr\'e model, we investigated a generalised Hamiltonian that also allows for quasiperiodic modulation of the hopping \cite{Hatsugai1990,Han1994,Liu2015}:
\begin{align}
    \nonumber H = - \sum_{n}  &\left[J + J\mu\cos \left(2\pi\beta\left(n+{\textstyle \frac12}\right)+\phi\right) \right]\\*
    &\times\left( a_n^\dagger a_{n+1} + \mathrm{H.c.}\right) 
    -J\lambda  \sum_{n} \cos(2\pi\beta n) a_n^\dagger a_n
    \label{eq:aubry_gen}
\end{align}
where $\mu$ is the dimensionless modulation amplitude of the hopping.  Remarkably, (\ref{eq:aubry_gen}) still has no mobility edges: localisation transitions occur  simultaneously in all eigenstates, similarly to the simple Aubry--Andr\'e case \cite{Han1994,Liu2015}. The boundary between extended and localised phases is given by
\begin{equation}
    \sum_\pm \sqrt{(\lambda/2)^2 \pm \lambda\mu \cos\phi + \mu^2} = 2
\end{equation}
for $\phi\neq0$, regardless of the value of $\beta$ \cite{Liu2015}. For $\phi=0$, the phase diagram consists of an extended ($\lambda<2;\mu<1$), a localised ($\lambda>2,2\mu$), and a critical phase ($2\mu>2,\lambda$) \cite{Liu2015,Han1994}. As examples, localisation transitions along the following paths were considered (see Fig.~\ref{fig:phases}):
\begin{subequations}
\label{eq:gen_instance}
\begin{align}
    \phi&=0, & \mu&=1/2, & \lambda&\approx2;
    \label{eq:gen_instance:o5}\\*
    \phi&=\pi/2, & 2\mu&=\lambda=\Lambda/\sqrt{2}, & \Lambda&\approx2.
\end{align}
\end{subequations}
Even though the hopping in these models is no longer uniform, $\Gamma$ was calculated  using the unchanged definition (\ref{eq:sff0}): it is an appropriate order parameter of the localisation transition regardless of normalisation.

Further to this generalised Aubry--Andr\'e model, we considered the continuum quasiperiodic Hamiltonian
\begin{equation}
    H = \frac{\hat{p}^2}{2m} + V_1 \cos^2(kx)+V_2\cos^2(\beta kx).
    \label{eq:twocos}
\end{equation}
Equation~\ref{eq:twocos} reproduces the Aubry--Andr\'e model in the limit $V_1\gg E_\mathrm{r}\gg V_2$ where the recoil energy,
\begin{equation*}
    E_\mathrm{r} = \frac{\hbar^2 k^2}{2m},
\end{equation*}
is the typical kinetic energy scale of the system. In addition to this limit, we studied the case of equal absolute lattice depths $V_1=V_2=VE_\mathrm{r}/2$. Periodic approximations to the Hamiltonian were implemented in momentum space and the curvature of the lowest band was calculated by exact diagonalisation using a formula adapted from (\ref{eq:sff}) \cite{Lieb2002,Roth2003}:
\begin{equation}
    \Gamma = \left.\frac{\pi^2}{E_\mathrm{r}}\frac{E_\Theta-E_0}{(\Theta/N)^2}\right|_{\Theta\to0} = \frac{m}{m_\mathrm{eff}}.
    \label{eq:sff_cont}
\end{equation}
A localisation transition was observed in the ground state of this model for all tested values of the incommensurate ratio $\beta$ at a $\beta$-dependent critical $V_c$. Unlike the generalised Aubry--Andr\'e model, however, its spectrum is unbounded, and several mobility edges appear in the spectrum of excited states. Nevertheless, we expect that the structure of the ground state has a hierarchical structure  similar to that discussed in Sec.~\ref{sec:intro}.

\begin{figure}
    \centering
    \includegraphics{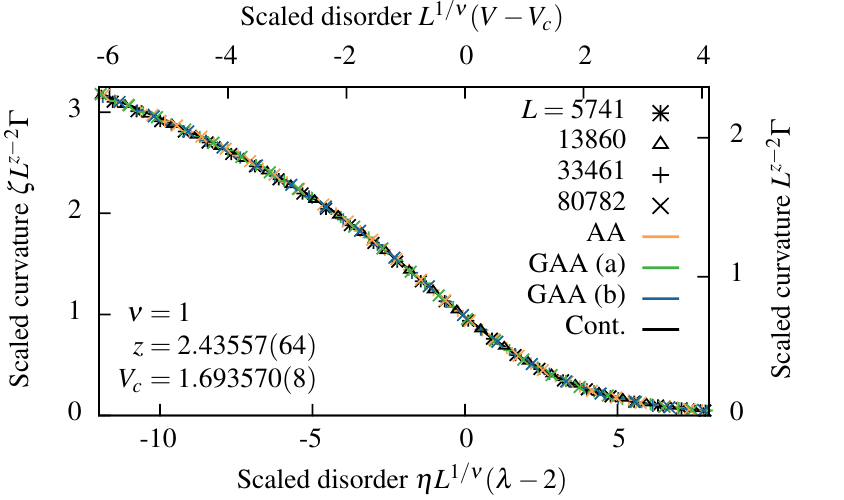}
    \caption{Finite-size scaling of  $\Gamma$ at the localisation transition of the Aubry--Andr\'e (AA) model, the generalised Aubry--Andr\'e (GAA) model with parameters (\ref{eq:gen_instance}) (bottom left axes), and the continuum model (\ref{eq:twocos}) with $V_1=V_2$ (top right axes) for $\beta = \sqrt2-1 = [0;\overline{2}]$. The scaling parameters for tight-binding models are $\eta_\mathrm{AA} = \zeta_\mathrm{AA}=1$; $\eta_a\approx1.155$, $\zeta_a\approx1.233$; $\eta_b\approx1.408$, $\zeta_b\approx1.203$. All models  share critical exponents and the data collapse onto the same scaling curve, suggesting they belong to the same universality class. }
    \label{fig:fss}
\end{figure}

To test this hypothesis, the curvature of the lowest band was computed for several rational approximations of $\beta=[0;\overline{2}]$ near the transition point of all these models. Since this continued fraction expansion is periodic, effective critical exponents $\nu$ and $z$ exist and can be determined using a finite-size scaling method \cite{Cestari2011,fss}. For a homogeneous system near a localisation transition, the finite-size scaling hypothesis can be applied to (\ref{eq:sff_scaling}) to give
\begin{equation}
    \Gamma = L^{z-2} \Phi\big(L^{1/\nu} \delta \big),
    \label{eq:sff_fss}
\end{equation}
where $L$ is the size of the finite system, $\delta$ is the distance from the transition point [\emph{e.g.,}~$(\lambda-2)$ for the Aubry--Andr\'e model] and $\Phi(x)$ is a scaling function determined by the universality class \cite{fss,Cestari2011}. In such systems, all sufficiently large length scales are equivalent: taking $\Gamma(\delta)$ for several different system sizes, critical exponents can be found accurately as the ones resulting in the best collapse of the scaled curves on each other \cite{fss,Melchert2009}. For quasiperiodic models, (\ref{eq:sff_fss}) does not hold in general, but for $\beta$'s with periodic continued fraction expansions, $N_k$'s separated by a full period of the expansion correspond to the same $\beta_k$ and thus display the same emergent structure. Using these values of $N_k$ as system sizes or period lengths, (\ref{eq:sff_fss}) applies and fitting to it yields the average dynamical exponent discussed in Sec.~\ref{sec:sff:aubry}.

The result of such a  fit is shown in Fig.~\ref{fig:fss}  for the Aubry--Andr\'e model, the generalised models (\ref{eq:aubry_gen},~\ref{eq:gen_instance}) and the continuum model (\ref{eq:twocos}) with $\beta=[0;\overline{2}]$. The resulting critical exponents are the same as are the scaling curves apart from overall rescaling. This suggests strongly that both the generalised Aubry--Andr\'e transitions and the continuum quasicrystal belong to the same ground state universality class as the Aubry--Andr\'e model.

\begin{figure}
    \centering
    \includegraphics{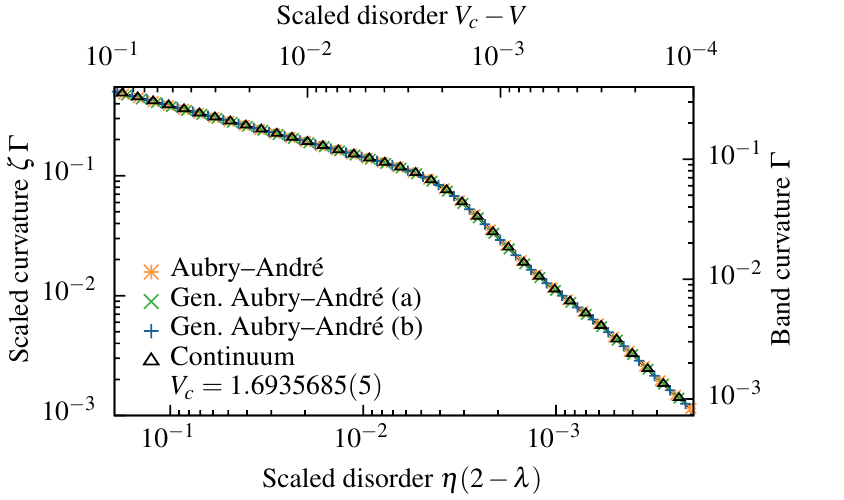}
    \caption{Curvature of the lowest band in the Aubry--Andr\'e model, the generalised Aubry--Andr\'e model with parameters (\ref{eq:gen_instance}) (bottom left axes), and the continuum model (\ref{eq:twocos}) with $V_1=V_2$ (top right axes) as a function of the reduced tuning parameter $\lambda-2$ and $V-V_\mathrm{c}$, respectively, for $\beta = [0;2,\dots,2,\overline{6}]$. The scaling parameters $\eta$ and $\zeta$ are the same as in Fig.~\ref{fig:fss}. Apart from overall rescaling, the scaling behaviour of all models are equivalent, suggesting they belong to the same universality class which, however, is not properly described by power-law scaling.}
    \label{fig:npl}
\end{figure}

For general $\beta$, however, the Aubry--Andr\'e phase transition has no well-defined critical exponents, and so the critical behaviour depends qualitatively on the correlation length. Therefore, such a universality class cannot be described in terms of critical exponents and finite-size scaling functions, only through the detailed dependence of observables on the length scale. To illustrate such universality, the curvature of the lowest band in all models was plotted in Fig.~\ref{fig:npl} as a function of the distance from the transition point. The curves can be collapsed on top of each other: points mapped onto each other correspond to an  equivalent correlation length. This notion of universality is markedly different from the conventional one based on the existence of a scaling regime in which the only effect of microscopic structure is to set critical exponents.

\section{Multifractal analysis}
\label{sec:multi}

The ground state dynamical exponent considered in Sec.~\ref{sec:sff} is a key quantity in quantum phase transitions, since at zero temperature, only the behaviour of the ground state is relevant. Unlike most quantum phase transitions, however, localisation transitions in the Aubry--Andr\'e model and its generalisation (\ref{eq:aubry_gen}) occur at the same point for all eigenstates \cite{Aubry1980,Han1994,Liu2015}, resulting in a fully singular continuous spectrum. A probe of the entire spectrum, as opposed to the ground state only, is also more relevant to experiments on Anderson and many-body localisation.

To explore the overall behaviour of the spectrum, we employed a multifractal scaling technique which yields statistics describing differences in the scaling behaviour at different parts of the spectrum. Furthermore, we demonstrate the connection between the structure of the spectrum and the resulting quantum dynamics by analysing the anomalous diffusion dynamics, a key experimental diagnostic, of the same models at criticality.

\subsection{Formulation}
\label{sec:multi:def}

Consider a periodic approximation $\beta=M_k/N_k$ of the incommensurate Hamiltonian. The singularity strength $\alpha_i$ of the $i$th subband is defined by
\begin{equation}
    \Delta_i \sim N_k^{-1/\alpha_i},
    \label{eq:multidef:alpha}
\end{equation}
where $\Delta_i$ is the width of the subband; by comparison to (\ref{eq:zdef:energy}), the ground state dynamical exponent is $1/\alpha$ for the lowest subband. For an incommensurate ratio with periodic continued fraction expansion, and hence a uniform scaling behaviour over different length scales, it is expected that the subbands of given singularity strength form a fully self-similar structure, the fractal dimension $f(\alpha)$ of which is given by \cite{Halsey1986,Tang1986}
\begin{equation}
    \Omega(\alpha) \sim \langle\Delta\rangle^{-f(\alpha)},
\end{equation}
where $\Omega(\alpha)\ud\alpha$ is the number of subbands with singularity strength between $\alpha$ and $\alpha+\ud\alpha$ and $\langle\Delta\rangle=N_k^{-1/\alpha}$ is a typical bandwidth of singularity strength $\alpha$. The function $f(\alpha)$ contains complete information about the scaling behaviour of the spectrum and is routinely used to characterise critical spectra of various systems \cite{Tang1986,Kohmoto1987,Han1994}. We note that the Hausdorff dimension of the entire spectrum is the maximum value of $f(\alpha)$ \cite{Tang1986}.

To accurately find $f(\alpha)$ numerically, we considered the scaling exponents $\tau_q$ defined through \cite{Tang1986}
\begin{equation}
    \sum_{i=1}^{N_k} \Delta_i^{-\tau_q} \sim N_k^{q}.
    \label{eq:multi:tau}
\end{equation}
This set of dimensions gives $f(\alpha)$ through the Legendre transform \cite{Halsey1986,Tang1986}
\begin{subequations}
\label{eq:multi:legendre}
\begin{align}
    \alpha &= \frac{\ud \tau_q}{\ud q};\\*
    f(\alpha) &=  q\alpha - \tau_q.
\end{align}
\end{subequations} 
It is now straightforward to show (see Appendix~\ref{app:multi}) that the slope of a straight line fit to
\begin{subequations}
\begin{equation}
    \aleph(N_k;\tau) = -\sum_i \mu_i^{(\tau)} \log \Delta_i
\end{equation}
and
\begin{equation}
    \phi(N_k;\tau) = -\sum_i \mu_i^{(\tau)} \log \mu_i^{(\tau)},
\end{equation}
respectively, as a function of $\log N_k$, where
\begin{equation}
    \mu_i^{(\tau)} = \frac{\Delta_i^{-\tau}}{\sum_j \Delta_j^{-\tau}},
\end{equation}
\end{subequations}
gives $\alpha^{-1}$ and $f/\alpha$ corresponding to a particular value of $\tau$; from this, the $f(\alpha)$ curve can be obtained parametrically.

\subsection{Results, universal multifractality}

\begin{figure}
    \centering
    \includegraphics{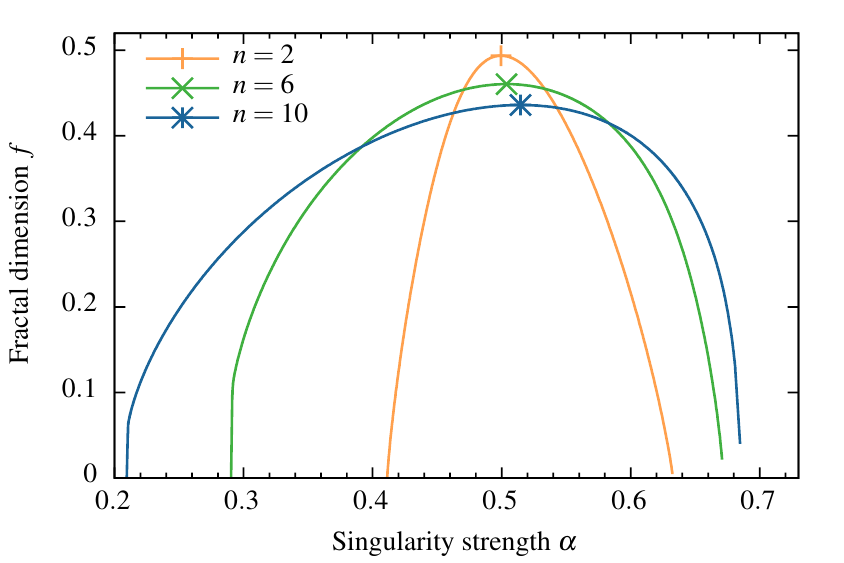}
    \caption{Multifractal dimensions $f(\alpha)$ for $\beta=[0;\overline{n}]$ ($n=2,6,10$) at the critical point of the Aubry--Andr\'e model. The smallest $\alpha$ in the spectrum coincides with the inverse of the ground state dynamical exponent, indicating the narrowest bands of the spectrum occur near the ground state. Symbols denote the peak of each curve: the most probable $\alpha$ is approximately 0.5 for small continued fraction terms, but significantly more for $n=10$. The $f(\alpha)$ curves of the critical generalised Aubry--Andr\'e Hamiltonian (\ref{eq:aubry_gen},\,\ref{eq:gen_instance}) are indistinguishably close to the ones plotted here.}
    \label{fig:multi}
\end{figure}

Multifractal analysis using the above formalism was carried out for $\beta=[0;\overline{n}]$ ($n=2,6,10$): the resulting $f(\alpha)$ curves for the Aubry--Andr\'e model are shown in Fig.~\ref{fig:multi}. $f(\alpha)$ is only defined on an interval $\alpha_\mathrm{min} \le \alpha \le \alpha_\mathrm{max}$ and $f(\alpha_\mathrm{min}) = f(\alpha_\mathrm{max}) = 0$: $\alpha_\mathrm{min,max}$ give the scaling exponents of the smallest and largest bandwidths of the system, respectively, but these represent a vanishing minority of all bands.  In fact, $\alpha_\mathrm{min}^{-1}$ equals the ground state dynamical exponent (\ref{eq:zdef:energy}) in all cases we considered. This suggests that the narrowest bands of the spectrum are near the bottom (and the top) of it and their scaling behaviour is atypical for the spectrum.

Localisation transitions in the generalised Aubry--Andr\'e model (\ref{eq:aubry_gen}) were also observed to occur simultaneously in all eigenstates \cite{Liu2015,Han1994}, giving rise to fully critical spectra at transitions. The multifractal dimensions $f(\alpha)$ at the transition points (\ref{eq:gen_instance}) were thus obtained using the same method. The $f(\alpha)$ curves for the simple and generalised Aubry--Andr\'e models are identical for a given $\beta$: this directly shows that the universality observed in the ground state also applies to the entire spectrum. For $\beta=[0;\overline{1}]$, the golden mean, and $\phi=0$, this behaviour was already known \cite{Han1994}. In this particular case, singular continuous spectra appear away from the localisation transition line as well (\emph{cf.}~Fig.~\ref{fig:phases}): in accordance with Ref.~\onlinecite{Han1994}, we found that the multifractal structure of these critical spectra is markedly different from the ones on the transition line (not shown). However, the existence of a critical region appears to be a peculiarity of the $\phi=0$ phase diagram \cite{Liu2015}, thus no universal features are expected of it.

It has been conjectured that the peak of the $f(\alpha)$ curve is at $\alpha^* = 1/2$ for all $\beta$, that is, the Hausdorff measure of the spectrum is dominated by bands scaling as $\Delta\sim N^{-2}$  \cite{Tang1986}. While this appears to be  the case for $\beta=[0;\overline{2}]$ and maybe for $[0;\overline{6}]$, it is certainly not  for $[0;\overline{10}]$ where $\alpha^*\approx0.515$ (the numerical error of $\alpha$ is at most $\approx 0.005$). Lower quality evidence for $[0;\overline{n}]$ with large $n$ suggests $\alpha$ increases further with $n$: the observation of Ref.~\onlinecite{Tang1986} appears to be a consequence of only using (more easily accessible) $\beta$'s with small continued fraction terms.

\subsection{Expansion of a wave packet}
\label{sec:expansion}

\begin{figure}
    \centering
    \includegraphics{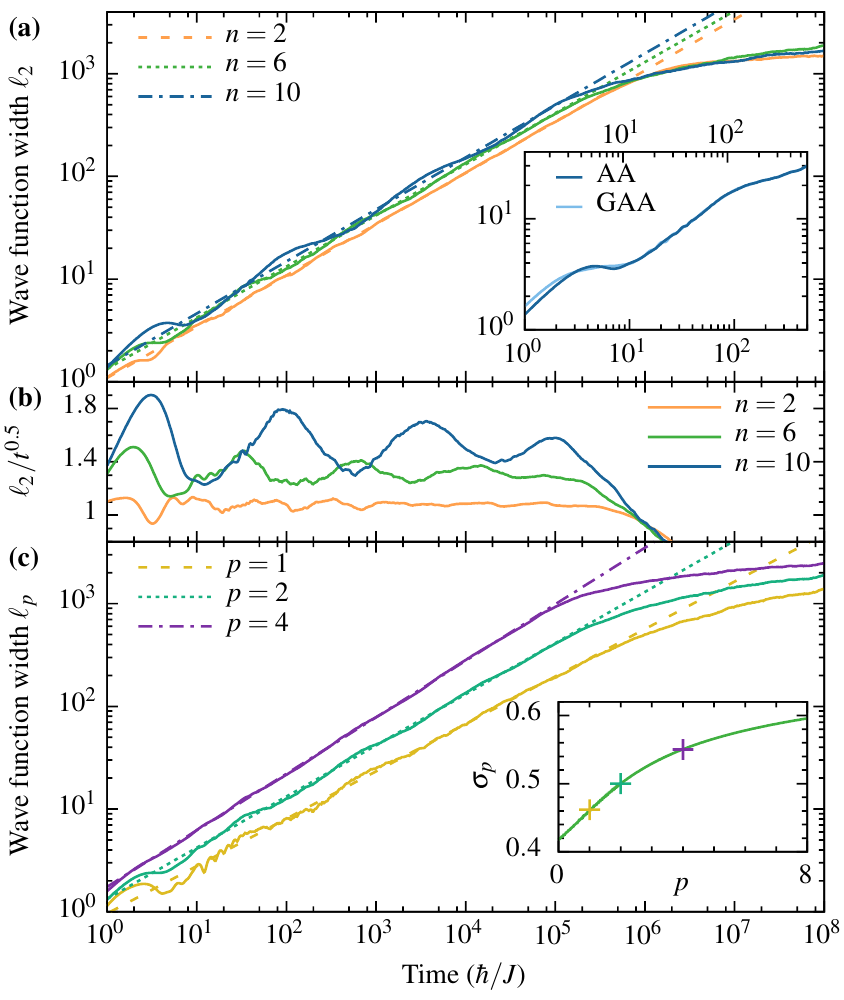}
    \caption{(a) RMS wave function width $\ell_2$ in the critical Aubry--Andr\'e model for rational approximations of $\beta=[0;\overline{n}]$ ($n=2,6,10$) as a function of time for a state initially localised on a single site, averaged over the initial site (solid lines). For all values of $\beta$, the expansion is well described by the power law $\ell_2\propto t^{1/2}$ (dashed and dotted lines). Convergence to a constant value at long times is a finite size effect.
    Inset: comparison of $\ell_2(t)$ for the simple ($\lambda=2$: AA, bottom time axis) and generalised [$(\lambda,\mu,\phi) = (2,1/2,0)$: GAA, top time axis] Aubry--Andr\'e models with $\beta=[0;\overline{10}]$. Except for very short times, the two curves are related by time dilation: $\ell_2^\mathrm{GAA}(t) = \ell_2^\mathrm{AA}(0.867t)$.
    (b) $\ell_2/t^{1/2}$ for the same expansions. As before, this ratio tends to a constant at long times, and the initial oscillations around this limit decay in time.
    (c) $\ell_p$ for $\beta=[0;\overline{6}]$ and $p=1,2,4$ in the same setup. For each $p$, $\ell_p$ increases as a power law, however, the critical exponents $\sigma_p$ depend on $p$ [$\sigma_1=0.4616(12)$, $\sigma_2=1/2$, $\sigma_4=0.5500(4)$]. Inset: comparison of $\sigma_p$ calculated from the multifractal spectrum using (\ref{eq:dynamics:exp}) (solid line) to the exponents obtained numerically (coloured crosses).}
    \label{fig:expansion}
\end{figure}

The multifractal dimensions $f(\alpha)$ contain full information on the scaling behaviour of the spectrum, and since the dynamics of a quantum system depends on differences between its energy levels, they capture the dynamical behaviour of the critical system. A straightforward example is the diffusion dynamics of an initially site-localised particle after a sudden quench onto the Aubry--Andr\'e Hamiltonian (\ref{eq:aubry_gen}). This expansion can be characterised through the evolution of the $p$th moment of the resulting quantum state:
\begin{align}
    \mu_p &= \langle |x-x_0|^p \rangle; &
    \ell_p &= \mu_p^{1/p},
\end{align}
where $x_0$ is the position where the wave function is initially localised and $p$ is an arbitrary positive real number. In a conventional critical system, $\ell_p\sim t^{1/z}$ because $t$ is a characteristic time scale corresponding to the length scale $\ell_p(t)$ \cite{Sachdev}. In this context, $\sigma=1/z$ is commonly referred to as the anomalous diffusion exponent.

Using exact diagonalisation, the time evolution of the initial state can be obtained directly from
\begin{equation}
    |\psi(t)\rangle = \sum_n |n\rangle e^{-iE_nt} \langle n | \psi(0)\rangle,
\end{equation}
where $|n\rangle$ are the eigenstates of the Hamiltonian with energy $E_n$: given $|\psi(t)\rangle$, $\mu_p$ can be calculated straightforwardly. 
As the details of the expansion dynamics will depend on the choice of initial state \cite{Varma2017}, we show in Fig.~\ref{fig:expansion}(a) the evolution of the rms width $\ell_2$ averaged over all initial sites $x_0$ for periodic approximations of $\beta=[0;\overline{n}]$ ($n=2,6,10$) in the Aubry--Andr\'e model. 
Apart from finite size effects, each expansion follows an approximate power law: fitting a power law to each plot resulted in a diffusion exponent $\sigma\approx0.5$ within the error of the fit. Similar behaviour has previously been found for other values of $\beta$ as well \cite{Hiramoto1988b}. On the other hand, $\sigma$ for a fixed value of $\beta$ does depend on $p$, as shown in Fig.~\ref{fig:expansion}(c) for $\beta=[0;\overline{6}]$ and $p=1,2,4$. This unusual behaviour is readily accessible by measuring higher moments of the diffused density distribution in typical sudden expansion experiments \cite{Schneider2012,Ronzheimer2013,Choi2016}.

In addition to the Aubry--Andr\'e model, $\ell_p(t)$ was calculated by the same method for the critical point $\lambda=2$, $\mu=1/2$, $\phi=0$ of the generalised Aubry--Andr\'e Hamiltonian: $\ell_2$ for $\beta=[0;\overline{10}]$ was plotted in the inset of Fig.~\ref{fig:expansion}(a) together with $\ell_2$ for the simple Aubry--Andr\'e model. The exponents of the approximate power laws were found to match, together with the structure of oscillations around it:
\begin{equation*}
    \ell_2^\mathrm{GAA}(t) = \ell_2^\mathrm{AA}(0.867t)
\end{equation*}
holds accurately for all but the shortest time scales.

\subsection{Connection between expansion dynamics and spectrum multifractality}

In order to connect the expansion dynamics in a critical tight-binding model to the multifractal properties of the spectrum, consider the Aubry--Andr\'e model with an arbitrary value of $\beta$ with periodic continued fraction expansion. Since the only natural length and time scales of the problem are the lattice spacing $a_0$ and the `hopping time' $\hbar/J$, $\mu_p$ depends on these scales as
\begin{equation}
    \mu_p(t;x_0,\beta,\lambda) = a_0^p m_p(Jt;x_0,\beta,\lambda),
\end{equation}
where $m_p$ is now a dimensionless function of dimensionless variables. To get an overall description of the critical dynamics, we set $\lambda=2$ and average over the position of the initial site:
\begin{equation}
    \overline{m}_p(Jt;\beta) = \lim_{N\to\infty}\frac1{2N+1}\sum_{x_0=-N}^N m_p(Jt;x_0,\beta,\lambda=2).
\end{equation}
Consider now the $k$th step of the renormalisation process outlined in Sec.~\ref{sec:spectrum:analytic}: the spectrum consists of $N_k$ critical subbands with incommensurate ratio $\beta_k$; let the effective hopping term in each be $J_i$ ($1\le i\le N_k)$. Provided the time $t$ is longer than the time scales corresponding to typical band gaps, interference between bands averages out, leaving
\begin{equation}
    m_p(Jt;x_0,\beta) \simeq  \sum_{i=1}^{N_k} \left|\left\langle x_0^{(i)} \middle| x_0\right\rangle\right|^2 N_k^p m_p\left(J_it;x_0^{(i)},\beta_k\right),
    \label{eq:dynamics:renorm}
\end{equation}
where $|x_0^{(i)}\rangle$ is the Wannier state of the $i$th subband living (among others) on site $x_0$;  the factor $N_k^p$ is due to the renormalisation of the lattice spacing. To average (\ref{eq:dynamics:renorm}) over lattice sites, we note that each renormalised band has one Wannier state per $N_k$ lattice sites and the sum of the overlap integrals $|\langle x_0^{(i)} | x_0\rangle|^2$ over all $x_0$ is 1 since the $|x_0\rangle$ form a basis. As a result, the overlap integrals average to $1/N_k$ for all lattice sites, and hence
\begin{equation}
    \overline{m}_p (Jt;\beta) \simeq N_k^{p-1} \sum_{i=1}^{N_k} \overline{m}_p(J_it;\beta_k).
    \label{eq:dynamics:renorm:avg}
\end{equation}

Now, consider those $k$ that correspond to full periods of the continued fraction expansion, that is, $\beta_k=\beta$. Assuming that the expansion is governed by a power law at long times,
\begin{equation}
    \overline{m}_p(Jt;\beta)\propto (Jt)^{p\sigma_p} \quad (Jt\to\infty),
    \label{eq:dynamics:assume}
\end{equation}
Eq.~\ref{eq:dynamics:renorm:avg} gives
\begin{align}
    \nonumber N_k^{p-1} \sum_{i=1}^{N_k} (J_it)^{p\sigma_p} &\simeq (Jt)^{p\sigma_p}\\
    \sum_{i=1}^{N_k} J_i^{p\sigma_p}& \propto \sum_{i=1}^{N_k} \Delta_i^{p\sigma_p} \propto N_k^{1-p}
    \label{eq:dynamics:scaling}
\end{align}
where $\Delta_i$ is the width of the $i$th subband for $\beta=M_k/N_k$, $4J_i$ in the unmodulated tight-binding approximation. In terms of the multifractal dimensions introduced in Sec.~\ref{sec:multi:def}, the anomalous diffusion exponents $\sigma_p$ are given by
\begin{equation}
    \sigma_p =-\frac{\tau_{1-p}}p.
    \label{eq:dynamics:exp}
\end{equation} 
In contrast to conventional diffusion dynamics, $\sigma_p$ now depends on $p$ and is not equal to the inverse of the ground state critical exponent. The only crucial assumption in deriving (\ref{eq:dynamics:exp}) is the self-similarity of the spectrum, therefore, we expect it to hold for the dynamics of other singular continuous spectra, \emph{e.g.,}\ the  Fibonacci quasicrystal \cite{Hiramoto1988a,Abe1987,Piechon1996}. In particular, as the spectra of all generalised Aubry--Andr\'e transition points are described by the same multifractal exponents, the $\sigma_p$ are universal too. The differences seen at very short times can be attributed to initial renormalisation steps required to attain a fixed point.

An interesting special case is that of $p=2$. There is strong numerical and analytical evidence \cite{Thouless1983,Thouless1990,Last1992,Tan1995} suggesting that for the Aubry--Andr\'e Hamiltonian with rational $\beta=M/N$, the sum of bandwidths scales as
\begin{equation}
    \lim_{N\to\infty} N\sum_{i=1}^N \Delta_i \approx 9.3299
    \label{eq:thouless}
\end{equation}
regardless of $M$. This implies that $\tau_{-1}=-1$ for any $\beta$: comparing with (\ref{eq:dynamics:exp}), we find that $\sigma_2 = 1/2$, as seen numerically in Fig.~\ref{fig:expansion}(a). Unlike diffusive systems, however, $\sigma=1/2$ here cannot be regarded as the consequence of a random walk between scatterers since $\sigma_p\neq1/2$ in general.

In Fig.~\ref{fig:expansion}(b), we note that oscillations around the approximate power law scaling of $\ell_p(t)$ decrease with time and become unnoticeable for sufficiently long times. The origin of this behaviour is clear from (\ref{eq:dynamics:renorm:avg}): for $Jt\gg1$, the expansion dynamics can be regarded as a superposition of the same dynamics at earlier times $J_it$. Since these $J_i$ range over several orders of magnitude for sufficiently large $N_k$, $\overline{m}(J_it)$ probes any short-time oscillations over several periods, thus averaging them out. That is, expansion length scales in different subbands can be very different, of which $\ell_p$ is only an average. This distinction becomes manifest in the expansion dynamics for $\beta$'s with aperiodic continued fraction expansions: while expansion dynamics at different length scales is different, at any particular time, these are averaged out, preventing the formation of clean crossovers similar to those seen in Fig.~\ref{fig:scaling} for a single sequence of subbands (namely, the ground state).

\section{Conclusion}
\label{sec:conclusion}

We have investigated the critical behaviour of the Aubry--Andr\'e model and other one-dimensional quasiperiodic systems near their localisation transitions. 
In particular, we considered the dependence of energy scales near the ground state, $\Delta E$, on the correlation length $\xi$. While the standard theory of phase transitions dictates that for large $\xi$, the system attains a scaling regime in which $\Delta E\propto \xi^{-z}$, we found that the critical behaviour is not described accurately by a power law on arbitrarily large length scales.

This is caused by the hierarchical structure of the critical spectrum of quasiperiodic models, captured by the continued fraction expansion of the irrational number $\beta=[0;n_1,n_2,\dots]$ describing their incommensurability. Each continued fraction term $n_k$ has associated with it a length scale $N_k$: scaling properties of the critical spectrum near this length scale were found to be fully determined by $n_k$.
Since the spectrum of a system near a phase transition  is sensitive to  spatial features on length scales up to the correlation length $\xi$, the critical behaviour of quasiperiodic models at $\xi\approx N_k$ will also be governed by $n_k$. As the sequence of these $n_k$ can be arbitrary and is controlled by the precise value of $\beta$, the dynamical exponent $z$ can typically not be defined for quasiperiodic models.  As an example, we found that for a wide class of $\beta$'s, $\Delta E$ tends to zero faster than any power of $\xi$, heralding critical behaviour qualitatively different from any conventional system.
Furthermore, the dependence of the critical behaviour on the incommensurate ratio is unusual: arbitrarily close values of $\beta$ can result in qualitatively different asymptotic behaviours very near the transition, as their continued fraction expansions eventually start to deviate.

Even though the localisation transition of one-dimensional quasiperiodic models cannot be described by power law relations, we find numerically that transitions in different models sharing the same value of $\beta$ display universal features. Instead of critical exponents, such universality classes are described  by the detailed dependence of observables such as $\Delta E$ on the correlation length. For models belonging to the same universality class, such functions can be scaled onto each other, similarly to finite-size scaling techniques for conventional phase transitions. The origin of such universality remains the identical behaviour under the renormalisation of length scales; the key difference is that quasiperiodic systems only admit a single sequence of discrete renormalisation steps that themselves depend on the length scale.

To complement studies of the ground state, we considered scaling properties of the entire spectrum on different length scales. For $\beta$'s with a periodic continued fraction expansion, the spectrum is expected to be self-similar at the Aubry--Andr\'e critical point: its structure was found to be a multifractal, and multifractal dimensions were calculated for several values of $\beta$. We also investigated the expansion dynamics of a localised wave packet and found that the evolution of the spread $\langle r^p\rangle^{1/p}$ of the wave function is described by a power law the exponent of which depends on $p$ and $\beta$. This is at odds with the behaviour of diffusive systems, where this exponent is $1/2$ for all $p$. Similarly to ground state properties, we again found universality between transition points of different quasiperiodic models in both their multifractal spectrum and expansion dynamics.

For the Aubry--Andr\'e model, we used a discrete renormalisation group protocol \cite{Suslov1982} to construct the critical spectrum and thus explicitly calculate the scaling of $\Delta E$ with $\xi$; non-power-law universality classes could be understood through the renormalisation behaviour of  other types of quasiperiodic models near phase transitions. 

Quasiperiodicity in higher dimensions leads to the emergence of arbitrarily large `microscopic' length scales the same way as in one dimension: this discrete large-scale structure is manifest in sharp diffraction peaks at progressively smaller momenta \cite{Shechtman1985,Mackay1982,Viebahn2018}. Therefore, it is reasonable to expect that phase transitions in such systems (including material quasicrystals) also display non-power-law behaviour. 
In general, quasiperiodic systems open the door to more complex large-scale behaviours, especially with interactions, which can show up, for instance, in increased quantum complexities \cite{Cubitt2015}, as novel universality classes for the many-body localisation transition  \cite{Khemani2017}, and in conjunction with their inherited topological features \cite{Kraus2012,Kraus2013}.

\section*{Acknowledgements}
We are grateful to Ehud Altman, Bartholomew Andrews, David Huse, and Austen Lamacraft for stimulating discussions and insights. This work was partly funded by the European Commision ERC starting grant QUASICRYSTAL and the EPSRC Programme Grant DesOEQ (EP/P009565/1).

\appendix
\section{WKB theory of tight-binding models}
\label{app:wkb}

In this appendix, we develop a semiclassical theory of tight-binding lattices with potentials slowly varying compared to the lattice spacing. The derivations presented here follow closely the standard derivations of WKB theory for an ordinary, quadratic dispersion relation \cite{landau3,landau9}. Since the period of the incommensurate modulation, $\beta^{-1}$ is large, this theory is applicable to the Aubry--Andr\'e model for the class of $\beta$'s considered, and can be used to accurately estimate the renormalised hopping and thus the critical exponents $\nu$ and $z$ \cite{Suslov1982}.

\subsection{Construction of the wave function}

We assume that the period of the modulating potential is very much larger than the lattice spacing. In this case, the discreteness of the wave function becomes irrelevant, and the Hamiltonian can be written as (the unit of length is the lattice spacing, $\hbar=1$)
\begin{equation}
    H = -2J\cos\hat{p} + V(\hat{x}).
    \label{eq:wkb:hamilton}
\end{equation}
where the nonquadratic dependence on $\hat{p}$ follows from the tight-binding dispersion relation. Due to this nonquadratic dispersion relation, the quasiclassical wave numbers depend differently on energy:
\begin{align}
    k(x)&= \arccos\left(\frac{V(x)-E}{2J}\right) ;
    \label{eq:wkb:k}\\*
    \kappa(x) = ik(x) &= \mathrm{arcosh} \left| \frac{E-V(x)}{2J}\right|,
    \label{eq:wkb:kappa}
\end{align}
Using $k(x)$, the Schr\"odinger's equation (\ref{eq:wkb:hamilton}) and the WKB ansatz can be written as
\begin{align}
    0 &= -(\cos \hat{p})\psi(x) + \cos k(x) \psi(x)
    \label{eq:wkb:schrodinger}\\
    \psi(x) &\approx A(x)  \exp\left(\pm i \int^x\! k(x')\ud x'\right) = A(x)\phi(x).
    \label{eq:wkb:wavefn}
\end{align}
where both $A(x)$ and $k(x)$ are assumed to vary slowly. Due to this slow variation, considering terms with a different number of derivatives amounts to separation of scales: in first order WKB approximation, only terms with zero or one derivatives are retained. The $n$th derivative of $\psi(x)$ is given by
\begin{align}
    \psi^{(n)}(x) &= A\phi^{(n)} + nA'\phi^{(n-1)} + O(A'')\quad \textrm{where}
    \label{eq:wkb:diff_psi}\\
    \nonumber \phi'(x) &= \pm ik \phi\\*
    \nonumber \phi''(x) &= (\pm ik)^2 \phi \pm ik'\phi\\*
    \nonumber \phi'''(x) &= (\pm ik)^3\phi + 3(\pm i)^2 kk'\phi \pm ik''\phi\\*
    \therefore \phi^{(n)}(x) &= (\pm ik)^n \phi + \binom{n}{2} (\pm i)^{n-1} k^{n-2}k'\phi +{O}(k'');
    \label{eq:wkb:diff_phi}
\end{align}
Eq.~\ref{eq:wkb:diff_phi} can be proved by induction. Combining (\ref{eq:wkb:diff_psi}) and (\ref{eq:wkb:diff_phi}) gives $\psi^{(n)}$ and $\hat{p}^n \psi$ as 
\begin{align}
    \nonumber \psi^{(n)} &= (\pm ik)^n A\phi + \binom{n}{2} (\pm ik)^{n-2} (\pm ik')A\phi \\*
    &\hspace{0.5in} +n (\pm ik)^{n-1} A'\phi;\\
    \nonumber \hat{p}^n \psi &= (\pm k)^n A\phi - i (\pm1)^{n-1} \binom{n}2 k^{n-2} k'A\phi \\*
    &\hspace{0.5in} -i n (\pm k)^{n-1} A'\phi.
\end{align}
Writing $\cos\hat{p}$ as a Taylor series, we finally obtain
\begin{align}
    \nonumber (\cos\hat{p})\psi &= \sum_{n=0}^\infty (-1)^n \frac{\hat{p}^{2n}\psi}{(2n)!}
     = \sum_{n=0}^\infty \bigg\{ (-1)^n \frac{k^{2n}}{(2n)!}\psi \mp\\*
    &\nonumber\quad i \left[\frac{(-1)^n}{2} \frac{k^{2n-2}}{(2n-2)!}k'A\phi + (-1)^n \frac{k^{2n-1}}{(2n-1)!}A'\phi \right] \bigg\}\\*
    &= \cos k\times\psi \pm i \left(\tfrac12\cos (k) k'A -\sin (k) A'\right)\phi.
\end{align}
Writing this into (\ref{eq:wkb:schrodinger}) yields
\begin{align}
    \nonumber \frac12\cos(k) k'A -\sin (k) A' &= 0\\
    A(x) &\propto \frac1{\sqrt{\sin k(x)}}.
\end{align}
Noting that the velocity of a classical particle moving under this Hamiltonian would be
\begin{equation}
    v = \dot{x} = \frac{\partial H}{\partial p} = 2J\sin p,
    \label{eq:wkb:speed}
\end{equation}
$A(x)$ can be interpreted as reproducing the classical probability of the particle being found at $x$, similarly to the amplitude in standard WKB theory \cite{landau3}.

The derivation above does not depend on $k(x)$ being real. At points with too large potentials, $k = i\kappa$ with $\kappa$ defined in (\ref{eq:wkb:kappa}), and the wave function (\ref{eq:wkb:wavefn}) becomes
\begin{equation}
    \psi(x) \propto \frac1{\sqrt{\sinh \kappa(x)}} \exp\left(-\int_{x_0}^x \kappa(x')\ud x'\right),
    \label{eq:wkb:decay}
\end{equation}
where the classical turning point $x_0$ is given by $E = V(x_0)-2$. At this turning point, $k=0$, and so the cosine dispersion may be replaced with a quadratic one: as a result, the Schr\"odinger's equation near the turning point reduces to the Airy equation. Solving this equation gives connection formulae equivalent to those in standard WKB theory:
\begin{align}
    \nonumber \frac{C}{2\sqrt{\sinh\kappa}} &\exp\left(-\int_{x_0}^x \kappa\ud x'\right) \\*
    &\longleftrightarrow \frac{C}{\sqrt{\sin k}} \cos\left(\int_{x_0}^x k\ud x -\frac\pi4\right). 
    \label{eq:wkb:conn}
\end{align}
The similarity of the connection formulae to standard WKB also means that the Bohr--Sommerfeld quantisation condition holds for this dispersion relation too:
\begin{equation}
    \int_{x_0}^{x_1} k(x)\ud x = \left(n+\frac12\right)\pi.
    \label{eq:wkb:bohr}
\end{equation}
We note that the region $E>V(x)+2$ is also inaccessible classically. There, $k(x)=i\kappa(x)+\pi$, corresponding to an exponentially decaying wave function changing signs at every lattice site. Eqs.~\ref{eq:wkb:decay}, \ref{eq:wkb:conn}, and~\ref{eq:wkb:bohr} generalise straightforwardly; we shall not discuss them in detail as they only become relevant near the top of the Aubry--Andr\'e spectrum.

Finally, we find the normalisation constant $C$ for a wave function living in a single potential minimum. Ignoring the exponentially decaying part, the normalisation requirement is
\begin{align*}
    1 &= \int_{x_0}^{x_1} \frac{C^2}{\sin p} \cos^2 \phi(x) \ud x \approx \int_{x_0}^{x_1} \frac{C^2}{\sin p} \frac12 \\*
    &= \frac{C^2}2 \int_{x_0}^{x_1} \frac{2J}{v}\ud x = JC^2  \int_{x=x_0}^{x_1} \ud t = JC^2 \frac{T}2= C^2 \frac{J\pi }{\omega}
\end{align*}
\begin{equation}
    C = \sqrt{\frac{\omega}{J\pi }} ,
    \label{eq:wkb:amplitude}
\end{equation}
where $\omega$ is the frequency of classical oscillations in the well.

\subsection{Hopping between neighbouring wells}

Consider a potential consisting of identical, centrosymmetric wells centred on $x=na$,  $n\in\mathbb{Z}$. If $a$ is large compared to the classically allowed region near the minimum of the potential, there is only appreciable hopping between neighbouring minima, and its value can accurately be estimated using WKB approximation. This calculation follows that of Ref.~\onlinecite{landau9} (\S{}55,~Problem~3) which solves the same problem for a quadratic dispersion.

Assuming that the overlap between wave functions $\Psi(x-an)$ living in neighbouring wells is small, each one can be treated as a Wannier function, that is, Bloch states are of the form
\begin{equation}
    \psi_k(x) = C\sum_{n=-\infty}^{\infty} e^{ikan}\Psi(x-an). 
    \label{eq:hop:bloch}
\end{equation}
The Schr\"odinger's equation for a single well and for the Bloch state are then 
\begin{subequations}
\begin{align}
    -J\psi_k(x-1) -J\psi_k(x+1) + [V(x)-\varepsilon_k]\psi_k(x) &= 0
    \label{eq:hop:schr_k}\\*
    -J\Psi(x-1) -J\Psi(x+1) + [V(x)-\varepsilon]\Psi(x) &= 0,
    \label{eq:hop:schr_site}
\end{align}
\end{subequations}
where $\varepsilon$ is the energy of a well state in isolation and $\varepsilon_k$ is the dispersion of the resulting band. Multiplying (\ref{eq:hop:schr_k}) by $\Psi(x)$, (\ref{eq:hop:schr_site}) by $\psi_k(x)$, subtracting and integrating from $x=-a/2$ to $a/2$ gives
\begin{widetext}
\begin{align}
    \nonumber (\varepsilon_k-\varepsilon)\psi_k(x)\Psi(x) &+J\left[\psi_k(x-1)\Psi(x) +\psi_k(x+1)\Psi(x) -\psi_k(x)\Psi(x-1) -\psi_k(x)\Psi(x+1)\right] = 0\\
    \nonumber (\varepsilon_k-\varepsilon) C &= -J \left[ \left(\int_{a/2-1}^{a/2} - \int_{-a/2-1}^{-a/2}\right) \Psi(x)\psi_k(x+1)\ud x + \left(\int_{-a/2}^{-a/2+1}-\int_{a/2}^{a/2+1}\right)\Psi(x)\psi_k(x-1)\ud x\right]\\*
    &= -J\left[\int_{\alpha-1/2}^{\alpha+1/2} \left\{\Psi\left(x-\frac12\right)\psi_k\left(x+\frac12\right) - \Psi\left(x+\frac12\right)\psi_k\left(x-\frac12\right)\right\}\right]^{a/2}_{\alpha=-a/2}.
    \label{eq:hop:disp_raw}
\end{align}

Consider the integral in brackets. At $\alpha=a/2$, only the $n=0$ and $n=1$ terms are relevant in (\ref{eq:hop:bloch}):
\begin{equation*}
    \psi_k(x) = C\left(\Psi(x)\pm \Psi(a-x)e^{ika}\right)
\end{equation*}
where the sign depends on whether the well eigenstate in question is even or odd. Substituting this form in the integral of (\ref{eq:hop:disp_raw}) gives
\begin{align*}
    \nonumber \int_{(a-1)/2}^{(a+1)/2} \cdots\ \ud x &= C \int_{(a-1)/2}^{(a+1)/2} \left[\Psi\left(x-\frac12\right)\Psi\left(x+\frac12\right) \pm \Psi\left(x-\frac12\right)\Psi\left(a-\frac12-x\right)e^{ika}\right.\\*
    \nonumber &\hspace{60pt} \left. -\Psi\left(x+\frac12\right)\Psi\left(x-\frac12\right) \mp \Psi\left(x+\frac12\right)\Psi\left(a+\frac12-x\right)e^{ika}\right]\ud x\\
    &= \mp Ce^{ika} \int_{-1/2}^{1/2} \left[\Psi\left(\frac{a+1}2+x\right)\Psi\left(\frac{a+1}2-x\right) -\Psi\left(\frac{a-1}2+x\right)\Psi\left(\frac{a-1}2-x\right) \right]\ud x.
\end{align*}
Similarly, for $\alpha=-a/2$, the $n=0$ and $n=-1$ terms yield
\begin{equation}
    \nonumber \int_{-(a+1)/2}^{-(a-1)/2} \cdots\ \ud x = \pm Ce^{-ika} \int_{-1/2}^{1/2} \left[\Psi\left(\frac{a+1}2+x\right)\Psi\left(\frac{a+1}2-x\right) -\Psi\left(\frac{a-1}2+x\right)\Psi\left(\frac{a-1}2-x\right) \right]\ud x,
\end{equation}
and hence by (\ref{eq:hop:disp_raw}), 
\begin{equation}
    \varepsilon_k-\varepsilon = \pm 2J\cos ka \int_{-1/2}^{1/2} \left[\Psi\left(\frac{a+1}2+x\right)\Psi\left(\frac{a+1}2-x\right) -\Psi\left(\frac{a-1}2+x\right)\Psi\left(\frac{a-1}2-x\right) \right]\ud x. \label{eq:hop:disp2}
\end{equation}
To evaluate each term of the integral in (\ref{eq:hop:disp2}), we employ a saddle point approximation to (\ref{eq:wkb:decay}): writing
\[\ln \sinh \kappa(\alpha+x) = \ln\sinh\kappa(\alpha) + rx + sx^2 + O(x^3),\]
we obtain
\begin{align}
    \nonumber \int_{-1/2}^{1/2} \Psi\left(\alpha+x\right)\Psi\left(\alpha-x\right)\ud x &\approx \Psi(\alpha)^2 \int_{-1/2}^{1/2} \exp\left(-\kappa x -\frac{\kappa'}2x^2 +\frac{r}2 x+\frac{s}2x^2\right) \exp\left(+\kappa x -\frac{\kappa'}2x^2 -\frac{r}2 x+\frac{s}2x^2\right)\ud x \\*
    &=\Psi(\alpha)^2 \int_{-1/2}^{1/2} e^{-(\kappa'-s)x^2}\ud x.
\end{align}
The Gaussian integral is only significantly different from 1 if the factor multiplying $x^2$ is $O(1)$; however, under the WKB approximation, $\kappa$ changes very slowly and so $\kappa', s\propto\kappa'' \ll 1$. That is, (\ref{eq:hop:disp2}) can be written as 
\begin{align*}
    \varepsilon_k-\varepsilon &\approx \pm 2J\cos ka \left[ \Psi^2\left(\frac{a+1}2\right) - \Psi^2\left(\frac{a-1}2\right)\right]
    \approx \pm 2J\cos ka\cdot \Psi^2\left(\frac{a}2\right) \left( e^{-\kappa} - e^{\kappa}\right) 
    = \mp 4J\cos ka \cdot \Psi^2\left(\frac{a}2\right) \sinh\kappa,
\end{align*}
where only the dominant variation in $\Psi(x)$ due to exponential decay was retained. Finally, substituting the wave function (\ref{eq:wkb:decay},\,\ref{eq:wkb:conn},\,\ref{eq:wkb:amplitude}) yields
\begin{align}
    \varepsilon_k-\varepsilon &\approx\mp \frac{\omega}\pi \exp\left(-2\int_{x_0}^{a/2} \kappa(x)\ud x\right)\cos ka.
\end{align}
That is, each well eigenstate broadens into a tight-binding type band with effective hopping term
\begin{equation}
    J' = \pm\frac{\omega}{2\pi} \exp\left(-2\int_{x_0}^{a/2} \kappa(x)\ud x\right).
    \label{eq:hop:final}
\end{equation}
\end{widetext}

\section{Renormalisation of $\lambda$}
\label{app:lambda}

From (\ref{eq:lambdat1}), the renormalised hopping is given by
\begin{equation}
    \lambda' = \lambda \frac{J'(4/\lambda,2E_0/\lambda;\beta)}{J'(\lambda,E_0;\beta)}
    \label{eq:lambda:renorm_def}
\end{equation}
where $J'(\lambda,E;\beta)$ is the hopping term (\ref{eq:hop:final}) for a band at energy $E$ in an Aubry--Andr\'e model with parameters $\lambda$ and $\beta$. Substituting (\ref{eq:hop:final}) gives
\begin{equation}
    \lambda' =  \lambda\frac{T}{T'} \exp\left(2\int_{x_0}^{1/(2\beta)}\kappa(x)\ud x - 2\int^{1/(2\beta)}_{x_0'}\kappa'(x)\ud x\right)
    \label{eq:lambda:form1}
\end{equation}
where $T$ is the classical period of oscillation around the minimum and $\kappa(x)$ is the imaginary wave vector (\ref{eq:wkb:kappa}); primes denote quantities of the dual model. For brevity, we write $\varepsilon = -E/J$.

\subsection{Relation of $T$ and $T'$}

From the classical velocity--momentum relation (\ref{eq:wkb:speed}),
\begin{align}
    \nonumber T &= \oint \frac{\ud x}{\dot{x}} = 4\int_0^{x_0} \frac{\ud x}{2J\sin k(x)} \\*
    \nonumber &= \frac4J \int_0^{x_0} \frac{\ud x}{\sqrt{4 - \big(\varepsilon-\lambda\cos(2\pi\beta x)\big)^2}} \\*
    \nonumber &= \frac{4}{2J\pi\beta } \int_0^{\alpha_0} \frac{\ud \alpha}{\sqrt{4-(\varepsilon-\lambda\cos\alpha)^2}} \\*
    & = \frac{4}{2J\pi\beta } \int_{\varepsilon-2}^\lambda \frac{\ud y}{\sqrt{\lambda^2-y^2}\sqrt{4-(\varepsilon-y)^2}},
    \label{eq:lambda:time}
\end{align}
where $\alpha=2\pi\beta x$ is the phase of the modulating potential and $y = \lambda \cos\alpha$. Very similarly, the classical period of the dual is 
\begin{align*}
    \nonumber T' &= \frac4{2J\pi\beta } \int_0^{\alpha_0'} \frac{\ud \alpha}{\sqrt{4-\left(\frac2\lambda \varepsilon -\frac4\lambda\cos\alpha\right)^2}}\\*
    &=\frac{2\lambda}{2J\pi\beta } \int_{\varepsilon-\lambda}^2 \frac{\ud y}{\sqrt{4-y^2}\sqrt{\lambda^2-(\varepsilon-y)^2}},
\end{align*}
where now $y = 2\cos\alpha$. The two integrals can be turned into each other by changing $y$ into $y'=\varepsilon-y$, therefore they are equal: (\ref{eq:lambda:form1}) becomes
\begin{equation}
    \lambda' = 2  \exp\left(2\int_{x_0}^{1/2\beta}\kappa(x)\ud x - 2\int^{1/2\beta}_{x_0'}\kappa'(x)\ud x\right).
    \label{eq:lambda:form2}
\end{equation}

\subsection{Evaluating the integrals $\int\!\kappa\,\ud x$}

\begin{figure}
    \centering
    \includegraphics{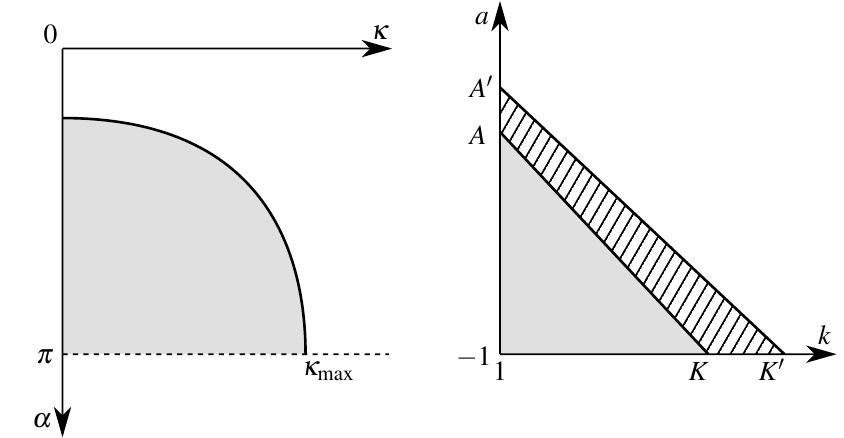}
    \caption{The integration domain of (\ref{eq:lambda:integrals1}) and (\ref{eq:lambda:cv}) after the change of variables $a = \cos\alpha$, $k=\cosh\kappa$ for $\lambda<2$. In the latter case, the integration domains of $I$ (gray) and $I'$ (striped and gray) are both right triangles and since $A<A'$, $K<K'$, the integration domain of $I-I'$ (striped) is a convex quadrilateral.}
    \label{fig:lambda:domain}
\end{figure}

To evaluate (\ref{eq:lambda:form2}), we first rewrite the integrals in terms of the phase $\alpha=2\pi\beta x$:
\begin{align}
    \nonumber\lambda' &= 2\exp\left(\frac{2}{2\pi\beta} \int_{\alpha_0}^\pi \kappa(\alpha)\ud\alpha - \frac{2}{2\pi\beta}\int_{\alpha_0'}^\pi \kappa'(\alpha)\ud\alpha\right) \\*
    &= 2\exp\left(\frac{I-I'}{\pi\beta}\right),
    \label{eq:lambda:form3}
\end{align}
where $I,I'$ are integrals independent of $\beta$, defined as
\begin{subequations}
\label{eq:lambda:integrals}
\begin{align}
    I &= \int_{\alpha_0}^\pi \mathrm{arcosh}\left(\frac{\varepsilon-\lambda\cos\alpha}2\right) \ud\alpha & \alpha_0 &= \arccos\left(\frac{\varepsilon-2}\lambda\right)
    \label{eq:lambda:integrals1}\\*
    I'&= \int_{\alpha_0'}^\pi \mathrm{arcosh}\left(\frac{ \varepsilon-2\cos\alpha}\lambda \right) \ud\alpha & \alpha_0'  &=\arccos\left(\frac{\varepsilon-\lambda}2\right).
\end{align}
\end{subequations}
These integrals can be thought of as the area in $(\alpha,\kappa)$ space bounded by $2\cosh\kappa+\lambda\cos\alpha=\varepsilon$ and $\lambda\cosh\kappa+2\cos\alpha=\varepsilon$, respectively. Introducing the variables $a = \cos\alpha$, $k=\cosh\kappa$, the area integrals can be rewritten as
\begin{equation}
    I,I' = \iint \frac{\ud a\,\ud k}{\sqrt{(1-a^2)(k^2-1)}};
    \label{eq:lambda:cv}
\end{equation}
the integration areas are bounded by the lines $a=-1$, $k=1$, and $2k+\lambda a = \varepsilon$ (for $I$) or $\lambda k + 2a =\varepsilon$ (for $I'$; see Fig.~\ref{fig:lambda:domain}). It follows that $I-I'$ entering (\ref{eq:lambda:form3}) is the integral of the same integrand over the difference of the two domains. Since for $\lambda<2$, 
\begin{align*}
    A' = \frac{\varepsilon-\lambda}2 &> \frac{\varepsilon-2}\lambda = A; &K' = \frac{\varepsilon+2}\lambda &> \frac{\varepsilon+\lambda}2 = K
\end{align*}
(and vice versa for $\lambda < 2$), this area difference is a quadrilateral bounded by all four lines bounding the triangles (see Fig.~\ref{fig:lambda:domain}).

For simplicity, we assume that $\lambda$ is infinitesimally close to 2: $\lambda=2+\eta, |\eta|\ll1$. In this case, the difference quadrilateral is infinitesimally thin: slicing it along lines of constant $a$ gives the integral
\begin{align}
    I-I' \simeq \int_{-1}^{A_0} \frac{\ud a [k(a)-k'(a)]}{\sqrt{1-a^2}\sqrt{k_0(a)^2-1}}.
    \label{eq:lambda:i}
\end{align}
In writing (\ref{eq:lambda:i}), we have ignored the variation of $k$ across one slice in the denominator, and replaced it with $k_0(a)$ corresponding to $\lambda=2$: this introduces first order corrections to the denominator which, since $k-k'$ is first order in $\eta$, can be ignored. Now,
\begin{equation*}
    k(a)-k(a') = \frac{\varepsilon-\lambda a}2 - \frac{\varepsilon -2a}\lambda \simeq \left(\frac\varepsilon4-a\right)\eta
\end{equation*}
\begin{align}
    \nonumber \therefore I-I'&\simeq \eta \int_{-1}^{(\varepsilon-2)/2} \frac{\ud a\left(\frac\varepsilon4-a\right)}{\sqrt{1-a^2}\sqrt{\left(\frac\varepsilon2-a\right)^2-1}} \\*
    \nonumber&= \eta \int_{-e+1}^{e+1} \frac{x\,\ud x}{\sqrt{\big(1-(e-x)^2\big)\big((e+x)^2-1\big)}}\\*
    \nonumber&= \frac{\eta}2\int_{(1-e)^2}^{(1+e)^2} \frac{\ud(x^2)}{\sqrt{4e^2-\big(x^2-(1+e^2)\big)^2}}\\
    I-I'&\simeq \frac{\pi\eta}2.
\end{align}
Writing this into (\ref{eq:lambda:i}) gives
\begin{equation}
    \lambda' \simeq 2e^{\eta/(2\beta)} \simeq 2+\frac{\eta}\beta;
\end{equation}
that is, the reduced tuning parameter $\lambda-2$ increases by a factor of $\beta^{-1}$ on rescaling.

It is possible to evaluate $I-I'$ for an arbitrary value of $\lambda$. We omit the derivation due to its length and report that
\begin{equation}
    I-I' = \pi\log\frac{\lambda}2 \implies \lambda' = 2\left(\frac\lambda2\right)^{1/\beta}
\end{equation}
as stated in Sec.~\ref{sec:spectrum:analytic}.

\section{Renormalisation of the hopping, the dynamical exponent}
\label{app:z}

\begin{figure}
    \centering
    \includegraphics{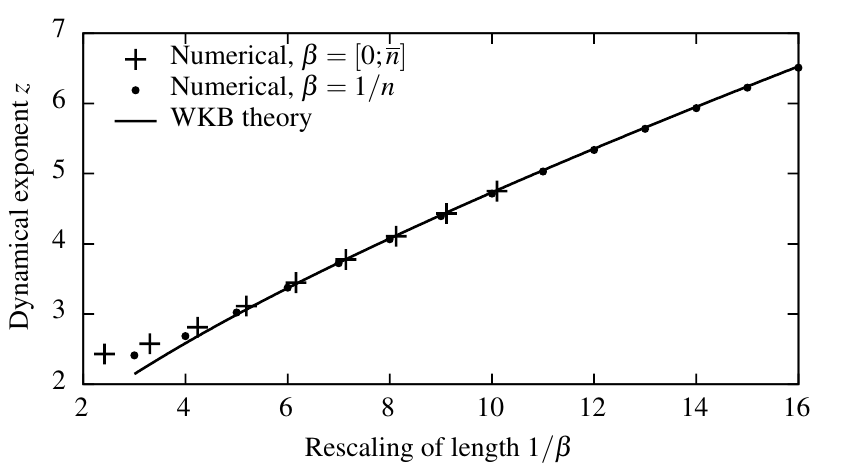}
    \caption{Comparison of effective dynamical exponents given by WKB theory with exact diagonalisation results for $\beta = [0;\overline{n}]$ and $\beta = 1/n$. For $\beta = [0;\overline{n}]$, $\beta_n = \beta$ at all RG steps and hence the dynamical exponent given by finite-size scaling is applicable to all steps. For $\beta=1/n$, the renormalisation of $J$ in a single RG step was obtained from the width of the lowest band. }
    \label{fig:z}
\end{figure}

To estimate the ground state dynamical exponent corresponding to a particular length scale, we consider the definition (\ref{eq:z_local}),
\begin{equation}
    z_k = \frac{\log \big(\Delta E(N_{k+1})/\Delta E(N_k)\big)}{\log \beta_k} \approx \frac{\log(J_{k+1}/J_k)}{\log\beta_k},
    \label{eq:z_local:def}
\end{equation}
at $\lambda=2$; since the renormalisation of $J$ in one RG step only depends on $\beta$ in that step, we anticipate that $z_k$ only depends on $\beta_k$.
We first consider the Bohr--Sommerfeld quantisation condition (\ref{eq:wkb:bohr}) for $\beta\ll1$: in terms of the phase $\alpha=2\pi\beta x$,
\begin{equation}
    \oint k(\alpha)\ud\alpha = 2(2n+1)\pi^2\beta,
    \label{eq:z_local:bohr}
\end{equation}
where the integrand is given by $2\cos k +2\cos\alpha = -E/J =\varepsilon$. For small values of $k$ and $\alpha$, both cosines can be approximated as quadratics: the contour of the area integral becomes approximately a circle, and thus
\begin{align}
    \nonumber \oint k(\alpha)\ud\alpha &\simeq \pi\alpha_0^2 \simeq \pi(4-\varepsilon)\\*
     \varepsilon_n &\simeq 4-2(2n+1)\pi\beta.
\end{align}
In particular, $n=0$ in the ground state, and so $\varepsilon_0 \simeq 4-2\pi\beta$. The most important consequence of this is that the ground state energy in the limit $\beta\ll1$ is close to $-4J$ and thus most of the distance between two neighbouring minima is classically unaccessible.

Consider now the expression (\ref{eq:hop:final}) of the renormalised hopping. By the quadratic approximation introduced above, the classical motion around a minimum can be treated as harmonic; the frequency follows from the coefficients of $p^2$ and $x^2$ as $\omega\simeq 4J\pi\beta $. Similarly to (\ref{eq:lambda:form3}), $J'$ can now be written as
\begin{align}
    J' &\simeq 2J\beta  \exp\left(-\frac{I}{\pi\beta}\right)\\
    \therefore z(\beta) & \simeq 1 + \frac{I}{\pi\beta|\log\beta|}
\end{align}
where $I$ is given by (\ref{eq:lambda:integrals1}); it also depends on $\beta$ through the ground state energy. Since $\varepsilon\approx4$ for any small $\beta$, the leading order term in $z(\beta)$ can be obtained by assuming $\varepsilon=4$ and thus $\alpha_0=0$:
\begin{align}
    I &\simeq \int_0^\pi \mathrm{arcosh}(2-\cos\alpha)\ud\alpha \approx 3.6639\\
    z(\beta) &\simeq 1.1662 \frac{\beta^{-1}}{\log(\beta^{-1})}.
\end{align}
That is, the ground state dynamical exponent diverges as $\beta_k\to0$, as discussed in Sec.~\ref{sec:sff}. More accurate estimates can be obtained by numerically solving (\ref{eq:z_local:bohr}) for $\varepsilon$ and evaluating (\ref{eq:hop:final}) directly.

To provide a numerical check on this result, the ground state dynamical exponent was obtained by the finite-size scaling method outlined in Sec.~\ref{sec:sff:universal} for $\beta_n=[0;\overline{n}]$, $2\le n\le 10$. For these numbers, $\beta_k=\beta$ for all $k$, and so the average dynamical exponent yielded by the finite-size scaling procedure equals $z(\beta)$. In addition, $z(1/n)$ was estimated by calculating the lowest bandwidth for $\beta=1/n$ and equating it to $4J'$ in the first and only step of the RG procedure. The resulting critical exponents are plotted against $\beta^{-1}$ in Fig.~\ref{fig:z} together with the $z(\beta)$ curve predicted by WKB theory. The correspondence between numerical and analytic results improves with decreasing $\beta_n$, as expected from the underlying assumptions of the analytic theory.

\section{Numerical computation of $f(\alpha)$}
\label{app:multi}

Due to its definition (\ref{eq:multi:tau}), it is more straightforward to obtain $q$ for a given value of $\tau$ than the other way around. Therefore, we consider the alternative Legendre transform
\begin{subequations}
\label{eq:multi:legendre2}
\begin{align}
    \alpha^{-1} &= \frac{\ud q}{\ud \tau};\\
    f' = f(\alpha)/\alpha &= q - \tau\alpha^{-1}.
\end{align}
\end{subequations}
In principle, this Legendre transform  could now be obtained from $q(\tau)$, given by power law fitting to (\ref{eq:multi:tau}), numerically: however, taking derivatives numerically tends to introduce significant noise. To mitigate this, we perform the Legendre transform \emph{before} power law fitting, as suggested by Ref.~\onlinecite{Chhabra1989}. Equation~\ref{eq:multi:tau} is equivalent to
\begin{equation}
    q = \lim_{N_k\to\infty} \frac{\log \left(\sum_i \Delta_i^{-\tau}\right)}{\log N_k};
\end{equation}
writing this into (\ref{eq:multi:legendre2}) gives
\begin{align*}
    \alpha^{-1}(\tau) &= \frac{\ud}{\ud \tau} \lim_{N_k\to\infty} \frac{\log \left(\sum_i \Delta_i^{-\tau}\right)}{\log N_k}\\*
    &= \lim_{N_k\to\infty} \frac1{\log N_k} \frac{\ud}{\ud \tau}\log \left(\sum_i \Delta_i^{-\tau}\right)\\*
    &= -\lim_{N_k\to\infty} \frac1{\log N_k} \sum_i \frac{\Delta_i^{-\tau} \log\Delta_i}{\sum_j \Delta_j^{-\tau}};\\
    f'(\tau) &= q - \tau\alpha^{-1} \\*
    &= - \lim_{N_k\to\infty} \frac1{\log N_k} \left[\sum_i \frac{\Delta_i^{-\tau} \log\Delta_i^{-\tau}}{\sum_j \Delta_j^{-\tau}} - \log \left(\sum_i \Delta_i^{-\tau}\right)\right]\\*
    &= - \lim_{N_k\to\infty} \frac1{\log N_k} \sum_i \frac{\Delta_i^{-\tau}}{\sum_j \Delta_j^{-\tau}} \log\frac{\Delta_i^{-\tau}}{\sum_j \Delta_j^{-\tau}}.
\end{align*}
That is, $\alpha^{-1}(\tau)$ and $f'(\tau)$ are given by fitting a straight line to
\begin{subequations}
\begin{align}
    \aleph(N_k;\tau) &= -\sum_i \mu_i^{(\tau)} \log \Delta_i,\\*
    \phi(N_k;\tau) &= -\sum_i \mu_i^{(\tau)} \log \mu_i^{(\tau)},
\end{align}
respectively as a function of $\log N_k$, where
\begin{equation}
    \mu_i^{(\tau)} = \frac{\Delta_i^{-\tau}}{\sum_j \Delta_j^{-\tau}},
\end{equation}
\end{subequations}
as stated in Sec.~\ref{sec:multi:def}.

\bibliography{bib/intro,bib/landau,bib/localisation,bib/superfluid,bib/rg,bib/multifractal,bib/expansion,bib/models}

\end{document}